\documentclass[12pt]{article}
\usepackage{amsmath}
\usepackage{braket}
\usepackage{ascmac}
\usepackage{color}
\usepackage{slashed}
\usepackage{url}
\usepackage{mathrsfs}

\numberwithin{equation}{section}

\usepackage[top=25mm, bottom=30mm, left=30mm, right=30mm]{geometry}

\font\mybb=msbm10 at 12pt
\def\bb#1{\hbox{\mybb#1}}
\def\Z {\bb{Z}}
\def\R {\bb{R}}
\def\C {\bb{C}}

\font\mybbs=msbm10 at 10pt
\def\bbs#1{\hbox{\mybbs#1}}

\def\sC {\bbs{C}}

\newcommand{\wt}{\widetilde}
\newcommand{\wh}{\widehat}
\newcommand{\ol}{\overline}

\newcommand{\ra}{\rightarrow}

\newcommand{\nn}{\nonumber}
\newcommand{\del}{\partial}
\newcommand{\half}{\frac12}
\newcommand{\VEV}[1]{\left\langle #1\right\rangle}
\newcommand{\Dslash}{\slashed{D}}
\newcommand{\DLslash}{\slashed{D}_+}
\newcommand{\DRslash}{\slashed{D}_-}
\newcommand{\cA}{{\cal A}}
\newcommand{\cD}{{\cal D}}
\newcommand{\cF}{{\cal F}}
\newcommand{\cH}{{\cal H}}
\newcommand{\cI}{{\cal I}}
\newcommand{\cJ}{{\cal J}}
\newcommand{\cO}{{\cal O}}
\newcommand{\Tr}{\textrm{Tr}}
\newcommand{\tr}{\textrm{tr}}
\newcommand{\Str}{\textrm{Str}}
\newcommand{\ch}{\textrm{ch}}
\newcommand{\sgn}{\textrm{sgn}}
\newcommand{\Ind}{\textrm{Ind}}
\newcommand{\AD}[1]{$\ol{\mbox{D~\,}}\!\!\!#1$}
\newcommand{\Gcov}{G^{\textrm{cov}}}
\newcommand{\Strsym}{\textrm{Str}^\textrm{sym}}
\newcommand{\sD}{{\mathscr D}}
\newcommand{\sJ}{\!{\mathscr J}}
\newcommand{\sP}{{\mathscr P}}

\def\mat#1{\matt[#1]}
\def\matt[#1,#2,#3,#4]{\left(%
\begin{array}{cc} #1 & #2 \\ #3 & #4 \end{array} \right)}

\def\vector#1{\vecc[#1]}
\def\vecc[#1,#2]{\left(#1\atop #2 \right)}

\begin{document}

\begin{flushright}
\hfill{YITP-21-41}~~~~
\end{flushright}
\begin{center}
\vspace{2ex}
{\Large \textbf{
Anomaly and Superconnection
}}

\vspace*{5mm}
\textsc{Hayato Kanno}$^a$\footnote{e-mail:
 \texttt{hayato.kanno@yukawa.kyoto-u.ac.jp}}
~and~
\textsc{Shigeki Sugimoto}$^{a,b}$\footnote{e-mail:
 \texttt{sugimoto@yukawa.kyoto-u.ac.jp}}

\vspace*{4mm}

\hspace{-0.5cm}
\textit{{$^a$
Center for Gravitational Physics, Yukawa Institute for Theoretical
 Physics,\\ Kyoto University, Kyoto 606-8502, Japan\ ,
}}\\
\textit{{$^b$
Kavli Institute for the Physics and Mathematics of the Universe (WPI),\\
 The University of Tokyo, Kashiwanoha, Kashiwa 277-8583, Japan
}}\\
\end{center}

\vspace*{.5cm}
\begin{abstract}
We study anomalies of fermions with spacetime dependent mass.
Using Fujikawa's method, it is found that the anomalies associated
with the $U(N)_+\times U(N)_-$ chiral symmetry and $U(N)$ flavor symmetry
for even and odd dimensions, respectively, can be written in terms of
superconnections. In particular, the anomaly for a vector-like $U(1)$
symmetry is given by the Chern character of the superconnection
in both even and odd dimensional cases.
It is also argued that the non-Abelian anomaly for a system
in $D$-dimensional spacetime is characterized by a $(D+2)$-form
part of the Chern character of the superconnection which
generalizes the usual anomaly polynomial for the massless case.
These results enable us to analyze anomalies
in the systems with interfaces
and spacetime boundaries in a unified way.
Applications to index theorems, including Atiyah-Patodi-Singer
index theorem and Callias-type index theorem, are also discussed.
In addition, we give a natural string theory interpretation
of these results.

\end{abstract}

\newpage
\tableofcontents

\section{Introduction}

Quantum anomaly is one of the fascinating topics
in quantum field theory. It implies important constraints
to have a consistent gauge theory and provides powerful tools to
investigate non-perturbative properties of quantum field theory.
It has been used to discuss phase structures of strongly
coupled systems and give non-trivial
evidence of conjectured dualities.
Another interesting aspect of the anomaly
is its beautiful mathematical structures.
In particular, the relations between the anomalies
and various index theorems have attracted much attention
and have been vigorously studied by both physicists and mathematicians.

In this paper, we investigate perturbative anomalies in
the systems with $N$ Dirac fermions including spacetime dependent mass
as well as external gauge fields associated with
$U(N)_+\times U(N)_-$ chiral symmetry or $U(N)$ flavor symmetry
for even or odd dimensional cases, respectively.
The spacetime dependent mass is equivalent to an external scalar
field (Higgs field) that couples with the fermions through
the Yukawa coupling. Although the masses of the quarks and leptons
in nature are considered to be constant, spacetime dependent mass
naturally appear in the standard model and various other models
when the value of the Higgs field is not constant.
It also appears in hadron physics and condensed matter physics,
because the effective mass of fermions can vary depending on
some parameters of the environment, such as temperature, chemical
potentials, magnetic field, strength of the interaction, etc.,
which can be spacetime dependent.

Apart from possible applications to realistic systems,
the spacetime dependent mass can be used as
a theoretical tool to study quantum field theory.
For example, it can be regarded as an external source coupled
to a fermion bilinear operator.
In particular, although the $U(N)_+\times U(N)_-$ chiral symmetry is
explicitly broken to a subgroup when the mass is non-zero,
we can make the action invariant under the $U(N)_+\times U(N)_-$ gauge
transformation (\ref{chiral})
by promoting the mass to a spacetime dependent external field.
Then, we are allowed to discuss the anomaly for this symmetry even though
the mass is non-zero. In this sense, the spacetime dependent
mass plays a similar role as the external gauge field, with which
the action becomes gauge invariant.
Furthermore, it can be used to study chiral fermions
localized on an interface or fermions in a spacetime with boundaries.
When we make the mass very large except for some regions in spacetime,
the low energy modes are trapped in the regions with small masses
and effectively induces a system with boundaries. If the mass profile
has a zero locus of non-zero codimension, it represents an interface
defined by the mass. As we review in section \ref{interface},
it is possible to realize Weyl fermions localized
in such interfaces. This mechanism is widely used to construct
theories with chiral fermions in lattice gauge theories,
phenomenological models of elementary particles
with extra dimensions, etc.

In fact, the anomaly for the fermions with spacetime dependent
mass (Higgs field) was analyzed in the 80's
in \cite{Fujikawa:1983bg,Hu:1983ij}\footnote{
See also section 6.5.1 of \cite{Fujikawa:2004cx}.}.
The conclusion of these papers was that the mass
does not contribute to the anomaly at all.
This is true in the case that the mass is bounded and fixed
while the cut-off scale is sent to infinity.
However, as we will demonstrate, the mass dependence of the
anomaly survives when the mass is unbounded.
Remarkably, we will also find that the anomaly exists even for odd
dimensional cases, when the spacetime dependent mass is introduced.
Our discussion is closely related to
that of recent papers by Cordova et al.
\cite{Cordova:2019jnf,Cordova:2019uob}, in which
coupling constants including the masses are promoted to external
scalar fields, and the anomalies are extended to include them.
They analyzed the systems with massive fermions in
\cite{Cordova:2019jnf} and found that the space of masses
can be considered as a compact space with non-trivial topology
by including $|m|\ra\infty$, and anomalies in $D$-dimensional systems
are characterized by a $(D+2)$-form, which is a generalization of
the usual anomaly polynomial, involving differential forms on the space
of masses. This also shows that it is crucial to consider $|m|\ra\infty$
to have a non-trivial anomaly that involves the masses.

The main goal of the first half of this paper (section \ref{Derivation})
is to show that the anomaly
$(D+2)$-form as well as the anomaly associated with $U(1)_V$ symmetry
are given by the Chern character written in terms of the superconnection
introduced by Quillen in \cite{Quillen:1985vya}.
This was also suggested in \cite{Cordova:2019jnf}.
We will show this explicitly by using Fujikawa's method.
Our formulas  (\ref{superID}) and (\ref{U1superD}) can be used for
both even and odd dimensional cases, provided that the superconnection
of the even and odd types are used accordingly.

These results are probably not surprising for those who are familiar
with the Chern-Simons (CS) terms including the tachyon field in unstable
D-brane systems, which are written with the Chern character of the
superconnection.\cite{Kennedy:1999nn,Kraus:2000nj,Takayanagi:2000rz,
Alishahiha:2000du} As we will discuss in section \ref{string},
the systems with Dirac fermions in various dimensions can be realized on
a D-brane with unstable D9-branes. The mass of the fermion is proportional
to the value of the tachyon field and hence the spacetime dependent
mass can be naturally obtained by considering a varying tachyon field.
The anomaly of the fermions is supposed to be canceled by the
contribution from the CS-term. Therefore, string theory suggests that
the superconnection appears in the formulas of anomaly, which is indeed
what we find in the field theory analysis.

The rest of the paper (section \ref{app}) is devoted to the applications
of these formulas.
We consider the systems with interfaces and boundaries realized
by the spacetime dependent mass.
Most of the discussion there are consistency checks and demonstration
of our formulas (\ref{superID}) and (\ref{U1superD}).
We show in several explicit examples that some known results can be
consistently reproduced in a simple and unified way. The results
of section \ref{evenbdry} are new. In this section, a system with
a spacetime dependent boundary condition is considered and the anomalies
due to this boundary condition are obtained.

This paper is organized as follows.
We start with a brief review of the superconnection in section
\ref{review}.
In section \ref{Derivation}, we derive our main formulas
for the anomaly with spacetime dependent mass
using Fujikawa's method.
Applications of these formulas are given
in section \ref{app}. The cases with interfaces
and boundaries are studied in sections \ref{interface}
and \ref{bdry}, respectively, and implications to index
theorems are discussed in section \ref{appindex}.
The systems with spacetime dependent mass can be realized
in string theory and our results have natural interpretations
in string theory as explained in section \ref{string}.
Finally, in section \ref{conclusion}, we summarize our results
and make concluding remarks.

\section{Superconnection}
\label{review}

In this section, we briefly review the superconnection introduced
by Quillen in \cite{Quillen:1985vya} with physicist-friendly notations.
Our description here is not as general as that given in the
original paper, but restricted to the cases to be used in the
following sections. See, e.g., \cite{Quillen:1985vya,BGV}
for more general and mathematically rigorous descriptions.
A superconnection\footnote{
In this paper, the word ``superconnection'' is used for the field
$\cA$ rather than the covariant derivative $d+\cA$,
which is often used in mathematical literature.
} $\cA$ of the even type is a matrix-valued field composed
of $U(N)\times U(N)$ gauge fields $(A_+,A_-)$ and a bifundamental
scalar field $T$ as
\begin{eqnarray}
\cA=\mat{A_+,iT^\dag,iT,A_-}
=A_+e^++A_-e^-+iT^\dag\sigma^++iT\sigma^-\ ,
\label{cA}
\end{eqnarray}
where
\begin{eqnarray}
 e^+=\mat{1,0,0,0}\ ,~~~ e^-=\mat{0,0,0,1}\ ,~~~
 \sigma^+=\mat{0,1,0,0}\ ,~~~ \sigma^-=\mat{0,0,1,0}\ .
\end{eqnarray}
In our notation, the gauge fields $A_\pm=A_{\pm\mu}(x) dx^\mu$
are one-forms that take values in anti-Hermitian $N\times N$ matrices.
$\sigma^\pm$ in (\ref{cA}) and $dx^\mu$ are treated as fermions,
\textit{i.e.}, they anti-commute with each other in the products.
The field strength of the superconnection
is defined as\footnote{The products of differential forms are the
wedge product, though the symbol for the wedge product `$\wedge$'
are omitted.}
\begin{eqnarray}
\cF\equiv d\cA+\cA^2=\mat{F_+-T^\dag T,iDT^\dag,iDT,F_--TT^\dag}\ ,
\label{cF}
\end{eqnarray}
where
\begin{eqnarray}
&& F_\pm\equiv dA_\pm+A_\pm^2\ ,
\nn\\
&&DT\equiv dT+A_-T-TA_+\ ,~~~
DT^\dag\equiv dT^\dag+A_+T^\dag-T^\dag A_-\ .
\end{eqnarray}

The Chern character is defined as
\begin{eqnarray}
\ch(\cF)\equiv
\sum_{k\ge 0}\left(\frac{i}{2\pi}\right)^{k/2}
\left[\Str\left(e^{\cF}\right)\right]_k\ ,
\label{chF}
\end{eqnarray}
where
$[\cdots]_k$
denotes the $k$-form part of the differential form
in the square brackets and
`$\Str$' is the supertrace\footnote{
In some literature, the symbol `$\Str$' is used for the symmetrized
trace, which should not be confused with the supertrace in this paper.
For the symmetrized trace, we use $\Tr^\textrm{sym}$.}
defined by
\begin{eqnarray}
 \Str\mat{a,b,c,d}\equiv \Tr(a)-\Tr(d)\ .~~~(\mbox{even case})
\label{str}
\end{eqnarray}
Because of (\ref{str}), only the even form part
in (\ref{chF}) can be non-zero.

A useful formula for a one-parameter family of superconnections denoted as
$\cA_t$ with a parameter $t\in[0,1]$ is
\begin{eqnarray}
\Str\left(e^{\cF_1}\right)-\Str\left(e^{\cF_0}\right)
= d\left(
\int_0^1 dt\,\Str\left(e^{\cF_t}\del_t\cA_t\right)
\right)\ ,
\label{eF1-eF0}
\end{eqnarray}
where $\cF_t=d\cA_t+\cA_t^2$.
For $\cA_t=\cA|_{T\ra tT}=\cA_0+t{\cal T}$ with $\cA_0= A_+e^++A_-e^-$
and ${\cal T}=iT^\dag\sigma^++iT\sigma^-$, this formula implies
\begin{eqnarray}
\Str\left(e^{\cF}\right)=
\Tr(e^{F_+})-\Tr(e^{F_-})
+ d\left(
\int_0^1 dt\,\Str\left(e^{\cF_t}{\cal T}\right)
\right)\ .
\label{eF1-eF0_2}
\end{eqnarray}
Since $\Str(e^{\cF_t}{\cal T})$ is gauge invariant,
(\ref{eF1-eF0_2}) implies that $\ch(\cF)$ and $\ch(F_+)-\ch(F_-)$
are equivalent up to an exact form.
For a trivial bundle (or, in a local patch)
the formula (\ref{eF1-eF0}) with $\cA_t=t\cA$ implies\footnote{
When the gauge group is $U(N_+)\times U(N_-)$ with $N_+\ne N_-$,
the right hand side has an additional constant term $N_+-N_-$.
}
\begin{eqnarray}
 \Str\left(e^\cF\right) = d\left(
\int_0^1 dt\,\Str\left(e^{td\cA+t^2\cA^2}\cA\right)
\right)\ .
\end{eqnarray}
This implies that the Chern character can be expressed
locally as
\begin{eqnarray}
\ch(\cF)=d\Omega
\label{chFOmega}
\end{eqnarray}
where $\Omega$ is the Chern-Simons (CS) form given by
\begin{eqnarray}
\Omega=\sum_{k\ge 0}\left(\frac{i}{2\pi}\right)^{(k+1)/2}
\left[\int_0^1dt\,\Str\left(e^{td\cA+t^2\cA^2}\cA\right)\right]_k\ .
\label{Omega0}
\end{eqnarray}
This $\Omega$ is, in general, not gauge invariant.

The superconnection of the odd type is
given by (\ref{cA}) with the restrictions
$A_+=A_-$ and $T=T^\dag$:
\begin{eqnarray}
\cA=\mat{A,iT,iT,A}
=A\, 1_2+iT\sigma_1\ ,
\label{cA2}
\end{eqnarray}
where $1_2=e^++e^-$ is the unit matrix of size 2 and
$\sigma_1=\sigma^++\sigma^-=\left({0\,1\atop 1\,0}\right)$.
The field strength is
\begin{eqnarray}
\cF\equiv d\cA+\cA^2=\mat{F-T^2,iDT,iDT,F-T^2}
\label{cF2}
\end{eqnarray}
with $F\equiv dA+A^2$ and $DT\equiv dT+[A,T]$.

The supertrace for the odd case is defined as
\begin{eqnarray}
 \Str\mat{a,b,b,a}\equiv  \sqrt{2}\,i^{-3/2}\,
\Tr(b)\ .~~~(\mbox{odd case})\ .
\label{Str2}
\end{eqnarray}
The reason for putting the normalization factor $\sqrt{2}\,i^{-3/2}$
will become clear later.\footnote{The sign ambiguity
of $i^{-3/2}$ is compensated by that of the $i^{k/2}$ factor
in (\ref{chF}). Namely, the supertrace $\Str$ of the odd case
always appears in the combination $i^{k/2}\Str$ with odd $k$
in the anomaly, and
$i^{k/2}\Str\left(a\,b\atop b\,a\right)=\sqrt{2}\,i^{(k-3)/2}\Tr(b)$
has no ambiguity.}
We also define an analog of the Chern character for the odd case
by the same formula as above (\ref{chF}). In this case,
only the odd form part contributes.
The formulas (\ref{eF1-eF0})--(\ref{Omega0}) also hold for the odd case.
In particular, (\ref{eF1-eF0_2}) with $A_+=A_-$ and $T=T^\dag$ gives
\begin{eqnarray}
\Str\left(e^{\cF}\right)=
 d\left(
\int_0^1 dt\,\Str\left(e^{\cF_t}iT\sigma_1\right)
\right)\ ,
\label{exact}
\end{eqnarray}
where $\cF_t=(F-t^2T^2)1_2+it DT\sigma_1$.
Therefore, the Chern character can also be written as
\begin{eqnarray}
\ch(\cF)=d\Omega'\ ,
\label{chFOmega2}
\end{eqnarray}
where
\begin{eqnarray}
\Omega'=\sum_{k\ge 0}\left(\frac{i}{2\pi}\right)^{(k+1)/2}
\left[\int_0^1dt\,\Str\left(e^{\cF_t}iT\sigma_1
\right)\right]_k\ .
\end{eqnarray}
Unlike $\Omega$ in (\ref{Omega0}), this $\Omega'$ is gauge
invariant.

\section{Derivation of the anomaly}
\label{Derivation}

\subsection{Even dimensional cases}
\label{even}

\subsubsection{Massive fermions and chiral anomaly}
\label{even1}

In this section, we consider a system with $N$ Dirac fermions $\psi$
in a $D=2r$-dimensional flat Euclidean spacetime ($r\in\Z_{> 0}$).
We include external gauge fields $A=(A_+,A_-)$ associated
with $U(N)_+\times U(N)_-$ chiral symmetry and a spacetime dependent
mass $m$, which belongs to the bifundamental representation
of $U(N)_+\times U(N)_-$.\footnote{
Although we discuss $N$ Dirac fermions, it is easy to get the results
for $N_\pm$ positive/negative chirality Weyl fermions
by considering a $U(N_+)_+\times U(N_-)_-$ subgroup of
$U(N)_+\times U(N)_-$ with large enough $N$.
}
 The action is
\begin{eqnarray}
S=
\int d^{D}\!x\,
\left(\ol{\psi}_+\DLslash\psi_+
+\ol{\psi}_-\DRslash\psi_-
+\ol\psi_- m\psi_+
+\ol\psi_+ m^\dag \psi_-
\right)
=
\int d^{D}\!x\, \ol{\psi}\cD
\psi\ ,
\label{action}
\end{eqnarray}
where\footnote{This notation is useful for our purpose, but is not
a standard one. A more standard notation is obtained by replacing
$\ol\psi$ and $\cD$ with $\ol\psi\left({0\,1\atop 1\,0}\right)$
and $\left({0\,1\atop 1\,0}\right)\cD$, respectively.}
\begin{eqnarray}
&&\psi(x)\equiv\vector{\psi_+(x),\psi_-(x)}\ ,~~~
\ol\psi(x)\equiv\left(\ol\psi_+(x),\ol\psi_-(x)\right)\ ,
\end{eqnarray}
and
\begin{eqnarray}
\cD\equiv
\mat{\DLslash,m^\dag(x),m(x),\DRslash}\ ,~~~
\DLslash\equiv\sigma^{\mu\dag}(\del_\mu+A_{+\mu})\ ,~~~
\DRslash\equiv\sigma^{\mu}(\del_\mu+A_{-\mu})\ .
\label{cD}
\end{eqnarray}
$\sigma^\mu$ and $\sigma^{\mu\dag}$ ($\mu=1,2,\cdots,D$) are
$2^{r-1}\times 2^{r-1}$ matrices
satisfying
\begin{eqnarray}
\sigma^{\mu\dag}\sigma^\nu+\sigma^{\nu\dag}\sigma^\mu
=\sigma^\nu\sigma^{\mu\dag}+\sigma^\mu\sigma^{\nu\dag}=2\delta^{\mu\nu}\ ,
\end{eqnarray}
so that
\begin{eqnarray}
\gamma^\mu\equiv\mat{0,\sigma^\mu,\sigma^{\mu\dag},0}
\ , ~~~(\mu=1,2,\cdots,2r)
\label{gamma}
\end{eqnarray}
are $D$-dimensional gamma matrices in a chiral representation. We choose
a representation of $\gamma^\mu$ such that
\begin{eqnarray}
 \gamma^1\gamma^2\cdots\gamma^{2r}=i^{r}\mat{1_{2^{r-1}},0,0,-1_{2^{r-1}}}
\equiv i^{r}\gamma^{2r+1}
\label{gamma5}
\end{eqnarray}
is satisfied, where $\gamma^{2r+1}$ is the chirality operator.

The crucial point here is that
we allow the mass parameter $m$
to depend on the spacetime coordinate $x^\mu$ and regard it as
an external scalar field, which is sometimes called a Higgs field
in the literature, that plays a similar role as the external
gauge fields $A_+$ and $A_-$.
Then, the classical action is invariant under $U(N)_+\times U(N)_-$
chiral gauge transformation that acts on the external fields as well
as the dynamical fermions as
\begin{eqnarray}
&&
\psi_+\to U_+\psi_+\ ,~~~
\ol\psi_+\to \ol\psi_+U_+^{-1}\ ,~~~
\psi_-\to U_-\psi_-\ ,~~~
\ol\psi_-\to\ol\psi_-U_-^{-1}\ ,
\nn\\
&&
A_+\ra U_+A_+ U_+^{-1}+U_+dU_+^{-1}\ ,~~~
A_-\ra U_-A_- U_-^{-1}+U_-d U_-^{-1}\ ,
\nn\\
&&
m\to U_- m\, U_+^{-1}\ ,~~~
m^\dag \to U_+ m^\dag U_-^{-1}\ ,
\label{chiral}
\end{eqnarray}
with $(U_+(x),U_-(x))\in U(N)_+\times U(N)_-$.

As it is well-known, the chiral symmetry is anomalous
in quantum theory. In fact, when the external fields are
non-trivial, the partition function
\begin{eqnarray}
Z[A,m]\equiv e^{-\Gamma[A,m]}
\equiv \int [d\psi d\ol\psi]\,e^{-S(\psi,\ol\psi,A,m)}
\label{pf}
\end{eqnarray}
gets a non-trivial phase under the chiral gauge transformation (\ref{chiral}),
even though the action is invariant.

Let us briefly review
the explicit form of the anomaly for the massless case.
Under an infinitesimal chiral gauge transformation
($U_+=e^{-v_+}$, $U_-=e^{-v_-}$ with $v_+,v_-\ll 1$) with
\begin{eqnarray}
\delta_v A_+=dv_++[A_+,v_+]\ ,~~~
\delta_v A_-=dv_-+[A_-,v_-]\ ,
\end{eqnarray}
The effective action for the massless case
 $\Gamma[A]\equiv \Gamma[A,m=0]$ defined in (\ref{pf})
transforms as $\Gamma\ra\Gamma+\delta_v\Gamma$ with
\begin{eqnarray}
\delta_v\Gamma[A]=
\int I^1_{2r}(v,A)\ ,
\label{delGam}
\end{eqnarray}
where $I^1_{2r}(v,A)$ is a $2r$-form obtained as a solution
of the descent equations\footnote{See, e.g.,
\cite{Bertlmann:1996xk,Fujikawa:2004cx,Harvey:2005it,Bilal:2008qx}
for reviews of the anomalies.}
\begin{eqnarray}
dI^1_{2r}=\delta_v I^0_{2r+1}\ ,~~~dI^0_{2r+1}=I_{2r+2}
\label{descent}
\end{eqnarray}
with
\begin{eqnarray}
I_{2r+2}(A)
=-2\pi i\left[\ch(F_+)-\ch(F_-)\right]_{2r+2}\ .
\label{ch}
\end{eqnarray}
Here,  $[\cdots]_{2r+2}$
denotes the $(2r+2)$-form part of the differential form
in the square brackets and $\ch(F_\pm)=\Tr\left(e^{\frac{i}{2\pi}F_\pm}\right)$
is the Chern character.
$I_{2r+2}(A)$ is called the anomaly polynomial and
$I^0_{2r+1}(A)$ is the CS $(2r+1)$-form.\footnote{
In this paper, we consider a flat spacetime.
For curved spacetime, $\ch(F)$ should be replaced with
$\ch(F)\wh A(R)$, where $\wh A(R)$ is the $\wh A$-genus.
Explicit expressions for $I_{2r+1}^0(A)$ and $I_{2r}^1(A)$ are
\begin{align}
I^0_{2r+1}(A)&=
\left(\frac{i}{2\pi}\right)^{r}\frac{1}{r!}
\int_0^1dt\,\left(\Tr(A_+F_{t+}^{r})-\Tr(A_-F_{t-}^{r})\right)\ ,
\\
I^1_{2r}(v,A)&=
\left(\frac{i}{2\pi}\right)^{r}\frac{1}{(r-1)!}\int_0^1dt\,
(1-t)
\left(\Tr^\textrm{sym}\left(v_+ d(A_+F_{t+}^{r-1})\right)
-\Tr^\textrm{sym}\left(v_- d(A_-F_{t-}^{r-1})\right)
\right)
\ ,
\end{align}
up to closed forms and contribution from local counterterms,
where $F_{t\pm}\equiv t dA_\pm+t^2 A_\pm^2$ and $\Tr^\textrm{sym}$
stands for the symmetrized trace.}

As pioneered by Fujikawa in \cite{Fujikawa:1979ay,Fujikawa:1980eg},
the chiral anomaly (\ref{delGam}) can be understood as a consequence
of the fact that the path integral measure for the fermions is not
invariant under the chiral transformation (\ref{chiral}).
After a careful regularization, it can be shown that
the fermion path integral measure transforms as
\begin{eqnarray}
 [d\psi d\ol\psi]\ra [d\psi d\ol\psi]\,\cJ
\label{jacobian}
\end{eqnarray}
with the Jacobian $\cJ$ given by
\begin{eqnarray}
\log\cJ
=\int I^1_{2r}(v,A)
\label{jacobian1}
\end{eqnarray}
under the infinitesimal chiral transformation, reproducing
the result in (\ref{delGam}).

The form of the Jacobian $\cJ$ in (\ref{jacobian}) depends on the
regularization. In \cite{Fujikawa:1980eg,Fujikawa:1983bg},
a manifestly gauge covariant form of the anomaly with
\begin{eqnarray}
\log\cJ=
\int I^{1\,\textrm{cov}}_{2r}(v,A)
\ ,
\label{jacobian2}
\end{eqnarray}
where
\begin{eqnarray}
I^{1\,\textrm{cov}}_{2r}(v,A)
=\left(\frac{i}{2\pi}\right)^{r}
\frac{1}{r!}\left(
\Tr(v_+F_+^{r})-\Tr(v_-F_-^{r})\right)
\label{covanom}
\end{eqnarray}
is obtained with a covariant regularization.
(See section \ref{anom}.)
This form of the anomaly is called the covariant anomaly, while
(\ref{delGam}) is called the consistent anomaly.
Unlike the consistent anomaly, the covariant anomaly
does not satisfy the descent equations (\ref{descent})
and cannot be written as the gauge variation of a well-defined
effective action. The consistent and covariant anomalies are related
by the addition of a Bardeen-Zumino counterterm in the associated
currents.\cite{Bardeen:1984pm} (See Appendix \ref{app:concov}.)

We are particularly interested in the anomaly for the $U(1)_V$
transformation which corresponds to
$v_+=v_-=-i\alpha(x)\, 1_N$ with a function $\alpha(x)$
and the unit matrix $1_N$.\footnote{
More precisely, what we are concerned here is
the mixed anomaly between $U(1)_V$ and
$SU(N)_+\times SU(N)_-\times U(1)_A$.}
In this case (\ref{covanom}) is
\begin{eqnarray}
I_{2r}^{1\,\textrm{cov}}(-i\alpha,A)
=-i\alpha\left[\ch(F_+)-\ch(F_-)\right]_{2r}
=\frac{\alpha}{2\pi}I_{2r}(A)
\ .
\label{U1anom}
\end{eqnarray}

The main claim of this section is that, when the spacetime
dependent mass $m$ is turned on, the Chern character
$\ch(F_+)-\ch(F_-)$ appeared in (\ref{ch}) and (\ref{U1anom})
are replaced with the Chern character written  by
the superconnection (\ref{chF}). More explicitly,
the anomaly polynomial $I_{2r+2}(A)$ in (\ref{ch}), the covariant anomaly
$I^{1\,\textrm{cov}}_{2r}(v,A)$ in (\ref{covanom})
and the $U(1)_V$ anomaly $I^{1\,\textrm{cov}}_{2r}(v,A)$ in (\ref{U1anom})
are replaced with
\begin{eqnarray}
I_{2r+2}(A,\wt m)&=&-2\pi i\,[\ch(\cF)]_{2r+2}\ ,
\label{superI}
\\
I^{1\,\textrm{cov}}_{2r}(v,A,\wt m)
&=&\left(\frac{i}{2\pi}\right)^{r}
\left[
\Str\left(v\,e^{\cF}\right)\right]_{2r}\ ,
\label{covsuper}
\\
I^{1\,\textrm{cov}}_{2r}(-i\alpha,A,\wt m)&=&
-i\alpha\,[\ch(\cF)]_{2r}\ ,
\label{U1super}
\end{eqnarray}
respectively, where $v\equiv{\textrm{diag}}(v_+,v_-)$,
$\wt m\equiv m/\Lambda$ is the mass rescaled
by the cut-off $\Lambda$ (see (\ref{sum}) for the definition) and
\begin{eqnarray}
\cF=\mat{F_+-\wt m^\dag\wt m,iD\wt m^\dag,iD\wt m,F_--\wt m\wt m^\dag}
\label{superF}
\end{eqnarray}
is the field strength of the superconnection (\ref{cF})
with $T=\wt m$.
(\ref{superI}) is related to $I_{2r}^1(v,A,\wt m)$
that gives the consistent anomaly
\begin{eqnarray}
 \delta_v\Gamma[A,m]=\int I^1_{2r}(v,A,\wt m)
\label{delGam2}
\end{eqnarray}
by the descent equation (\ref{descent}).\footnote{
See section \ref{interface} for more on the use of
the anomaly $(D+2)$-form (\ref{superI}).}
Since $I_{2r+2}(A,\wt m)$ is not a polynomial of the field strength
$\cF$, we refer to it as an anomaly $(2r+2)$-form following \cite{Cordova:2019jnf}.
(\ref{covsuper}) is the covariant anomaly related to the Jacobian $\cJ$
defined with a covariant regularization adopted in section \ref{anom} by
\begin{eqnarray}
 \log \cJ=\int I^{1\,\textrm{cov}}_{2r}(v,A,\wt m)\ .
\label{logJ2}
\end{eqnarray}
(\ref{U1super}) is obtained from (\ref{covsuper}) by setting
$v_+=v_-=-i\alpha 1_N$.
In (\ref{delGam2}) and (\ref{logJ2}), we take $\Lambda\ra\infty$ limit
after the integration.

Note that when $m$ is bounded, $\wt m$ vanishes in the limit
$\Lambda\ra\infty$ and the $m$ dependence drops
out.\cite{Fujikawa:1983bg,Hu:1983ij}
However, there are some physically interesting systems
in which the mass is of the order of cut-off scale or unbounded,
and the $m$ dependence in the anomaly may survive.
For example, a system with a boundary can be realized
by setting $m\ra\infty$ in a region of the spacetime. Another interesting
example is a system with localized massless fermions on an interface
(defect) with mass of order cut-off scale in the bulk,
such as the domain-wall fermions used in lattice QCD \cite{Kaplan:1992bt}.
We will consider such examples in section \ref{app}.

Another related issue is that, as it was shown in \cite{Quillen:1985vya},
the de Rham cohomology class of (\ref{superI}) is independent of $\wt m$
because of the relation (\ref{eF1-eF0_2}), which would mean that the $m$
dependent part of
$(2r+2)$-form (\ref{superI}) does not contribute to the anomaly.
This is true in a compact spacetime. However, for an open space,
the $\wt m$ dependent part of the anomaly $(2r+2)$-form can give
a non-trivial element of the cohomology with
compact support.\footnote{
See \cite{Cordova:2019jnf} for more on this point.}
As we will discuss in section \ref{interface},
this non-trivial element is interpreted as the anomaly of the
fermions localized on the interfaces located around the zero
locus of the mass profile. The local counterterm that cancels this
anomaly is the contribution from the anomaly inflow.

In section \ref{anom}, we will explicitly show (\ref{covsuper})
and (\ref{U1super})
using Fujikawa's method, following the prescription given in
\cite{Fujikawa:1983bg}.
Our argument for (\ref{superI}) is more
indirect. This is suggested as a consequence of
the relation (\ref{U1anom})
between the non-Abelian anomaly in $2(r-1)$-dimensions characterized by
$I_{2r}(A)$ and the Abelian anomaly given by
$I_{2r}^{1\,\textrm{cov}}(-i\alpha,A)$ in $2r$-dimensions
\cite{Zumino:1983rz,AlvarezGaume:1983cs}
generalized to the cases with spacetime dependent mass.
This issue will be discussed in section \ref{ID+2}.

\subsubsection{Calculation of the Jacobian}
\label{anom}

In order to show (\ref{covsuper}) and (\ref{U1super}), we evaluate
the Jacobian $\cJ$ in (\ref{jacobian}) for the $U(N)_+\times U(N)_-$
transformation (\ref{chiral}). In the following, we demonstrate
the derivation of (\ref{U1super}) in detail focusing on
the $U(1)_V$ transformation that acts on the fermions as
\begin{eqnarray}
 \psi(x)\ra e^{i\alpha(x)}\psi(x)\ ,~~~
 \ol\psi(x)\ra e^{-i\alpha(x)}\ol\psi(x)\ ,
\label{U1V}
\end{eqnarray}
which is a special case of the transformation in (\ref{chiral}) with
$U_+=U_-=e^{i\alpha}1_N$.
The generalization to general $U(N)_+\times U(N)_-$ transformations
that leads to (\ref{covsuper}) is straightforward.

Following \cite{Fujikawa:1983bg},
we expand the fermion fields $\psi$ and $\ol\psi$ with respect to
the eigenfunctions of the Hermitian operators $\cD^\dag\cD$ and
$\cD\cD^\dag$, respectively.
Let $n_\phi$ and $n_\varphi$ be the number of zero modes
of $\cD^\dag\cD$ and $\cD\cD^\dag$, respectively, and choose the
eigenfunctions such that they satisfy the eigenequations\footnote{
Here, we have assumed that the spectra of $\cD^\dag\cD$
and $\cD\cD^\dag$ are discrete. Later, we will consider the cases
with non-compact spacetime. In such cases, the asymptotic behavior
of the mass and the gauge fields should be chosen appropriately
to have discrete spectra.
}
\begin{eqnarray}
&&\cD^\dag\cD\varphi_n(x)=\lambda_n^2\varphi_n(x)\ ,~~~
(n\in\{\,k-n_\varphi\,|\,k=1,2,3,\cdots\,\})\ ,
\label{eigeneq1}\\
&&\cD\cD^\dag\phi_n(x)=\lambda_n^2\phi_n(x)\ ,~~~
(n\in\{\,k-n_\phi\,|\,k=1,2,3,\cdots\,\})\ ,
\label{eigeneq2}
\end{eqnarray}
and the normalization conditions
\begin{eqnarray}
\int d^{D}\!x\,\varphi_m^{\dag}(x)\varphi_n(x)
=\delta_{m,n}\ ,~~~
\int d^{D}\!x\,\phi_m^{\dag}(x)\phi_n(x)
=\delta_{m,n}\ .
\label{norm}
\end{eqnarray}
Here, the eigenvalues of $\cD^\dag\cD$ and $\cD\cD^\dag$ are denoted as
$\lambda_n^2$, because they are non-negative and can be written
as the square of real numbers.\footnote{
Be aware that $\lambda_n$ is not the eigenvalue of $\cD$. $\cD$ is not
Hermitian and its eigenvalues are not real in general.}
Without loss of generality, we assume
 $\lambda_n=0$ for $n\le 0$ and
$0<\lambda_1\le\lambda_2\le\lambda_3\le \cdots$.
Note that the eigenvalues for (\ref{eigeneq1}) and (\ref{eigeneq2})
are the same, because the non-zero modes
 $\varphi_n$ and $\phi_n$ with $n>0$ are related by
\begin{eqnarray}
\phi_n(x)=\frac{1}{\lambda_n}\cD\varphi_n(x)\ ,~~~
\varphi_n(x)=\frac{1}{\lambda_n}\cD^\dag\phi_n(x)\ ,~~~(\mbox{for}~n>0)\ ,
\end{eqnarray}
up to phase.

Then, fermions $\psi(x)$ and $\ol\psi(x)$ can be expanded as
\begin{equation}
\psi(x)=\sum_na_n\varphi_n(x)\ ,~~~
\ol{\psi}(x)=\sum_n\ol{b}_n\phi_n^{\dag}(x)\ ,
\end{equation}
where $a_n$ and $\ol{b}_n$ are Grassmann odd coefficients,
and the action (\ref{action}) becomes
\begin{eqnarray}
S=\sum_n\lambda_n\ol{b}_na_n\ .
\end{eqnarray}
The fermion path integral measure is formally defined as
\begin{eqnarray}
 [d\psi d\ol\psi]=\prod_{x}d\psi(x)d\ol\psi(x)
=\det(\varphi_n(x))^{-1}\det(\phi_n^\dag(x))^{-1}\prod_n da_n\prod_m
d\ol b_m\ ,
\end{eqnarray}
where $\det(\varphi_n(x))^{-1}\det(\phi_n^\dag(x))^{-1}$ is the Jacobian
for the change of variables from $\{\psi(x),\ol\psi(x)\}$ to
$\{a_n,\ol b_n\}$.

Under the $U(1)_V$ transformation (\ref{U1V}),
$a_n$ and $\ol b_n$ transforms as
\begin{eqnarray}
a_n&\to& a_n'
\equiv \int d^{D}\!x\,
\varphi_n^{\dag}(x)e^{i\alpha(x)}\psi(x)\simeq
\sum_{m}\left(\delta_{m,n}
+i\int d^{D}\!x\,\varphi_n^{\dag}(x)\alpha(x)\varphi_m(x)\right)a_m\ ,
\nn\\
\ol b_n&\to&\ol b_n'
\equiv \int d^{D}\!x\,\ol\psi(x) e^{-i\alpha(x)}\phi_n(x)\simeq
\sum_{m}\ol b_m
\left(\delta_{m,n}
-i\int d^{D}\!x\,\phi_m^{\dag}(x)\alpha(x)\phi_n(x)\right)\ ,
\nn\\
\end{eqnarray}
where we have assumed $\alpha(x)\ll 1$.
Then, the Jacobian (\ref{jacobian}) is
\begin{eqnarray}
\log \cJ
=-i\int d^{D}\!x\,\alpha(x)\cI(x)\ ,
\label{Jacobian}
\end{eqnarray}
where
\begin{eqnarray}
\cI(x)\equiv \sum_n\left(
\varphi_n^\dag(x)\varphi_n(x)-\phi_n^\dag(x)\phi_n(x)
\right)\ .
\end{eqnarray}
$\cI(x)$ can be regularized by introducing a UV cut-off $\Lambda$ as
\begin{eqnarray}
\cI(x)&=&\lim_{\Lambda\ra\infty}\sum_n e^{-\frac{\lambda_n^2}{\Lambda^2}}
\left(
\varphi_n^\dag(x)\varphi_n(x)-\phi_n^\dag(x)\phi_n(x)
\right)\nn\\
&=&
\lim_{\Lambda\ra\infty}\sum_n
\left(
\varphi_n^\dag(x)e^{-\frac{1}{\Lambda^2}\cD^\dag\cD}\varphi_n(x)
-\phi_n^\dag(x)e^{-\frac{1}{\Lambda^2}\cD\cD^\dag}\phi_n(x)
\right)\nn\\
&=&
\lim_{\Lambda\ra\infty}\int\frac{d^{D}k}{(2\pi)^{D}}e^{-ikx}\,
\Tr_s\left(
e^{-\frac{1}{\Lambda^2}\cD^\dag\cD}
-e^{-\frac{1}{\Lambda^2}\cD\cD^\dag}
\right)e^{ikx}\ ,
\label{sum}
\end{eqnarray}
where $\Tr_s$ is the trace over both flavor and spinor
indices.
The cut-off $\Lambda$ will be sent to infinity at the end of the
calculation.\footnote{
The explicit form of the anomaly actually depends on the choice
of the regularization scheme. We adopt this heat kernel
regularization in a covariant form.
}

To evaluate (\ref{sum}), note that $\cD^\dag\cD$ and $\cD\cD^\dag$ are
written as
\begin{eqnarray}
\cD^\dag\cD=-D_\mu^2-\Lambda^2\wh\cF\ ,~~~
\cD\cD^\dag=-D_\mu^2-\Lambda^2\wh\cF'\ ,
\label{DD}
\end{eqnarray}
where
\begin{eqnarray}
 D_\mu=\mat{\del_\mu+A_{+\mu},0,0,\del_\mu+A_{-\mu}}\ ,
\end{eqnarray}
and
\begin{eqnarray}
\wh\cF
&=&
\mat{\frac{1}{2\Lambda^2}
\sigma^\mu\sigma^{\nu\dag}F_{+\mu\nu}
-\wt m^\dag\wt m,
\frac{1}{\Lambda}\sigma^\mu D_\mu\wt m^\dag,
\frac{1}{\Lambda}\sigma^{\mu\dag} D_\mu\wt m,
\frac{1}{2\Lambda^2}
\sigma^{\mu\dag}\sigma^{\nu}F_{-\mu\nu}-\wt m\wt m^\dag}\ ,
\\
\wh\cF'
&=&
\mat{\frac{1}{2\Lambda^2}
\sigma^{\mu\dag}\sigma^{\nu}F_{+\mu\nu}
-\wt m^\dag\wt m,
-\frac{1}{\Lambda}\sigma^{\mu\dag} D_\mu\wt m^\dag,
-\frac{1}{\Lambda}\sigma^{\mu} D_\mu\wt m,
\frac{1}{2\Lambda^2}
\sigma^{\mu}\sigma^{\nu\dag}F_{-\mu\nu}-\wt m\wt m^\dag}
\end{eqnarray}
with $\wt m\equiv m/\Lambda$.
Then, (\ref{sum}) becomes
\begin{eqnarray}
\cI(x)
&=&\lim_{\Lambda\ra\infty}\int\frac{d^{D}k}{(2\pi)^{D}}
\Tr_s\left(
e^{\frac{1}{\Lambda^2}(ik_\mu+D_\mu)^2+\wh\cF}
-e^{\frac{1}{\Lambda^2}(ik_\mu+D_\mu)^2+\wh\cF'}
\right)
\nn\\
&=&\lim_{\Lambda\ra\infty}\Lambda^{D}
\int\frac{d^{D}\wt k}{(2\pi)^{D}} e^{-\wt k_\mu^2}\,
\Tr_s\left(
e^{\frac{1}{\Lambda^2}D_\mu^2+\frac{2i}{\Lambda}\wt k^\mu D_\mu+\wh\cF}
-e^{\frac{1}{\Lambda^2}D_\mu^2+\frac{2i}{\Lambda}\wt k^\mu D_\mu+\wh\cF'}
\right)\ ,
\label{cI}
\end{eqnarray}
where $\wt k_\mu\equiv k_\mu/\Lambda$.
Using the formula
\begin{eqnarray}
\tr\left(
\sigma^{\mu_1}\sigma^{\mu_2\dag}\cdots
\sigma^{\mu_{2k-1}}\sigma^{\mu_{2k}\dag}
-
\sigma^{\mu_1\dag}\sigma^{\mu_2}\cdots
\sigma^{\mu_{2k-1}\dag}\sigma^{\mu_{2k}}
\right)
=\left\{
\begin{array}{ll}
~0&(k<r)\\
(2i)^r\epsilon^{\mu_1\cdots\mu_{2r}}&(k=r)
\end{array}
\right.
\ ,
\end{eqnarray}
where $\epsilon^{\mu_1\cdots\mu_{2r}}$ is the Levi-Civita symbol
with $\epsilon^{1,2,\cdots,D}=1$,
and assuming that
the gauge field, $\wt m$ and $\wt k_\mu$ as well as their derivatives
are all of $\cO(1)$ in the $1/\Lambda$ expansion,
\footnote{
In section \ref{app}, we consider the cases with $m$ being a linear function
of $x^\mu$. One may wonder whether $\wt m$ can be regard as an
$\cO(1)$ parameter, even though $\wt m$ diverges at $|x|\ra\infty$.
In that case, our treatment here can be understood as the evaluation of
the $\Lambda\ra\infty$ limit of the integration
$\int d^Dx\,\alpha(x)\cI(x)$ by using rescaled coordinates
$\wt x^\mu= x^\mu/\Lambda$.
}
it is easy to verify
\begin{eqnarray}
\cI(x)=\lim_{\Lambda\ra\infty}\Lambda^{D}
\int\frac{d^{D}\wt k}{(2\pi)^{D}} e^{-\wt k_\mu^2}\,
\Tr_s\left(
e^{\wh\cF}-e^{\wh\cF'}
\right)
=
\lim_{\Lambda\ra\infty}
\frac{\Lambda^{D}}{2^D\pi^{D/2}}
\Tr_s\left(e^{\wh\cF}-e^{\wh\cF'}\right)\ ,
\label{cI2}
\end{eqnarray}
and
\begin{eqnarray}
\Tr_s\left(e^{\wh\cF}-e^{\wh\cF'}\right)d^{2r}\!x
=\Lambda^{-2r}(2i)^{r}\,\left[\Str\left(e^{\cF}\right)\right]_{2r}
+\cO(\Lambda^{-2r-1})\ ,
\end{eqnarray}
where $d^{2r}\!x=dx^1\cdots dx^{2r}$ and $\cF$ is
the superconnection defined in (\ref{superF}).
Neglecting the $\cO(\Lambda^{-1})$ terms,
this implies\footnote{
This formula (in the $\Lambda\ra\infty$ limit with $\wt m$ kept fixed)
corresponds to the local index theorem proved in
\cite{Getzler}. See also \cite{Kahle}.}
\begin{eqnarray}
 \cI(x)d^{2r}\!x=\left(\frac{i}{2\pi}\right)^r
\left[\Str\left(e^{\cF}\right)\right]_{2r}
=
\left[\ch(\cF)\right]_{2r}\ ,
\label{Iint}
\end{eqnarray}
and hence we obtain
\begin{eqnarray}
 \log\cJ=-i
\int \alpha(x)
\left[\ch(\cF)\right]_{D}\ ,
\label{U1super2}
\end{eqnarray}
which is the desired result (\ref{U1super}).

In section \ref{app}, we consider the cases
with $A_+=A_-$ and the
mass given by a scalar matrix as
\begin{eqnarray}
m=\mu(x)1_N\ ,
\label{scalarm1}
\end{eqnarray}
where $\mu(x)$ is a complex function and
$1_N$ is the unit matrix of size $N$. In this case,
we have
\begin{eqnarray}
\ch(\cF)=
\frac{i}{2\pi}d\wt\mu^\dag d\wt\mu\,
e^{-|\wt\mu|^2}
\ch(F)
\end{eqnarray}
with $F\equiv F_+=F_-$ and $\wt\mu\equiv\mu/\Lambda$,
and the Jacobian (\ref{U1super2}) becomes
\begin{eqnarray}
\log\cJ=\frac{1}{2\pi}
\int d\wt\mu^\dag d\wt\mu\,
e^{-|\wt\mu|^2} \alpha(x)
\left[\ch(F)\right]_{D-2}\ .
\label{logJscalar1}
\end{eqnarray}

\subsubsection{Anomaly $(D+2)$-form}
\label{ID+2}

In this subsection, we give a simple derivation of the anomaly
$(D+2)$-form (\ref{superI}) using the result (\ref{U1super})
for the $U(1)_V$ anomaly. Although the description here is for
the even dimensional case, the argument is applicable to the odd
dimensional case as well.

We decompose the $U(N)_+\times U(N)_-$
gauge fields into the $U(1)_V$ gauge field $V$ and the rest,
and write the Chern character as
\begin{eqnarray}
 \ch(\cF)=e^{\frac{i}{2\pi}f^V}\ch(\cF_0)\ ,
\label{chFchF}
\end{eqnarray}
where $f^V\equiv dV$ is the field strength of the $U(1)_V$ gauge field
and $\cF_0\equiv\cF|_{f^V=0}=\cF-f^V 1_{2N}$.
First, we try to show (\ref{superI}) for the case with $f^V=0$.
To this end, let us consider the $U(1)_V$ anomaly
(\ref{U1super}) with $f^V=0$ in a $(D+2)$-dimensional system:
\begin{eqnarray}
I^{1\,\textrm{cov}}_{D+2}(-i\alpha,A,\wt m)|_{f^V=0}=
-i\alpha[\ch(\cF_0)]_{D+2}\ .
\label{U1super3}
\end{eqnarray}
Note that for this component of the anomaly, there is no difference
between the covariant and consistent anomalies.\footnote{We thank
Y. Tanizaki for the discussion on this point.} (See Appendix \ref{WZcon}.)
The anomaly $(D+4)$-form for the $(D+2)$-dimensional
system that reproduces (\ref{U1super3}) via the descent equations
(\ref{descent}) is
\begin{eqnarray}
f^V[\ch(\cF_0)]_{D+2}\ .
\label{D+4form}
\end{eqnarray}

Now, consider a $(D+2)$-dimensional spacetime of the form
$S^2\times M_D$, where $M_D$ is a $D$-dimensional manifold.
We assume that $f^V$ has a flux with $\int_{S^2}f^V=-2\pi i$
and $\cF_0$ is independent
of the coordinates on $S^2$. In this case, each fermion in the
$(D+2)$-dimensional system has one zero mode on $S^2$
and hence we get a $D$-dimensional system with $N$ Dirac
fermions in the limit that the radius of the $S^2$
becomes zero. The anomaly $(D+2)$-form for this $D$-dimensional
system is given by integrating (\ref{D+4form}) over $S^2$,
yielding
\begin{eqnarray}
I_{D+2}(A,\wt m)|_{f^V=0}=\int_{S^2} f^V[\ch(\cF_0)]_{D+2}=
-2\pi i\,[\ch(\cF_0)]_{D+2}\ ,
\end{eqnarray}
which is (\ref{superI}) for the $f^V=0$ case.

The $f^V$ dependence of the anomaly
$(D+2)$-form can be easily recovered by replacing $\cF_0$ with $\cF$,
which completes the derivation of (\ref{superI}).

\subsection{Odd dimensional cases}

\subsubsection{Anomaly in odd dimensions}
\label{Ainodd}

In this section, we consider a system with $N$ Dirac fermions $\psi$ in
a $D=(2r+1)$-dimensional flat Euclidean spacetime ($r\in\Z_{\ge 0}$).
The flavor symmetry is $U(N)$ and the associated external gauge field
is denoted as $A$. We include a spacetime dependent mass $m$,
which is a Hermitian matrix of size $N$ and belongs to the adjoint
representation of $U(N)$. The action is
\begin{eqnarray}
S=\int d^{D}\!x\, \ol\psi\left(\Dslash+m\right)\psi
=\int d^{D}\!x\, \ol\psi \cD\psi\ ,
\label{Sodd}
\end{eqnarray}
where
\begin{eqnarray}
\Dslash\equiv\gamma^\mu(\del_\mu+A_\mu)\ ,~~~
\cD\equiv\Dslash+m\ ,
\label{Dslashodd}
\end{eqnarray}
and $\gamma^\mu$ ($\mu=1,2,\cdots,2r+1$) are gamma matrices
satisfying $\gamma^{\mu\dag}=\gamma^\mu$ and
 $\{\gamma^{\mu},\gamma^{\nu}\}=2\delta^{\mu\nu}$.
For explicit computation, we choose $\gamma^\mu$
to be of the form (\ref{gamma}) for $\mu=1,\cdots,2r$
and $\gamma^{2r+1}$ in (\ref{gamma5}) for $\mu=2r+1$.
This action is invariant under the $U(N)$ flavor symmetry:
\begin{eqnarray}
 \psi\ra U\psi\ ,~~~\ol\psi\ra\ol\psi U^{-1}\ ,~~~
A\ra UAU^{-1}+UdU^{-1}\ ,~~~m\ra UmU^{-1}
\end{eqnarray}
with $U(x)\in U(N)$.

Our claim is that the formulas analogous to
(\ref{superI}), (\ref{covsuper}) and (\ref{U1super}):
\begin{eqnarray}
I_{D+2}(A,\wt m)&=&-2\pi i\,[\ch(\cF)]_{D+2}\ ,
\label{superID}
\\
I^{1\,\textrm{cov}}_D(v,A,\wt m)
&=&\left(\frac{i}{2\pi}\right)^{D/2}
\left[
\Str\left(v\,e^{\cF}\right)\right]_{D}\ ,
\label{covsuperD}
\\
I^{1\,\textrm{cov}}_{D}(-i\alpha,A,\wt m)&=&
-i\alpha\,[\ch(\cF)]_{D}\ ,
\label{U1superD}
\end{eqnarray}
hold even for the odd dimensional cases, using
the odd dimensional analog of the Chern character
(\ref{chF}) defined by the supertrace for the odd case
(\ref{Str2}).
Unlike the even dimensional cases discussed in section \ref{even},
both (\ref{superID}) and (\ref{U1superD})
vanish when the mass $m$ vanishes. The anomaly appears
only when $m$ is turned on.

We will show in section \ref{anom2} that the formula
(\ref{U1super2}) for the $U(1)_V$ transformation (\ref{U1V})
also holds for the odd dimensional cases
by examining the Jacobian of the fermion path
integral measure using Fujikawa's method.
This implies (\ref{U1superD}). The derivation can be easily
generalized to (\ref{covsuperD}). (\ref{superID}) follows from
(\ref{U1superD}) by an indirect argument given in section \ref{ID+2}.

The meaning of (\ref{superID}) is somewhat more ambiguous,
because, for odd $D$, we can find a gauge invariant
$(D+1)$-form $I^0_{D+1}(A,\wt m)$ satisfying
$I_{D+2}(A,\wt m)=dI_{D+1}^0(A,\wt m)$
(See (\ref{chFOmega2}).)
Then, the odd dimensional analogue of the descent equations
(\ref{descent}):
\begin{eqnarray}
dI^1_{D}=\delta_v I^0_{D+1}\ ,~~~dI^0_{D+1}=I_{D+2}
\label{descentD}
\end{eqnarray}
would imply that the anomaly $I^1_D$ simply vanishes.
However, as we will see in section \ref{kink},
$I^0_{D+1}(A,\wt m)$ is non-vanishing at infinity in our examples
with non-trivial interfaces and $I_{D+2}$
can be a non-trivial element of the cohomology with
compact support.\footnote{\label{foot}
A similar statement holds for the
mass dependent part of the Chern character $I_{2r}(A,\wt m)$
for the even dimensional case, as mentioned in section \ref{even1}
and demonstrated in section \ref{vortex}.
}
We will argue that the anomaly of the fermions on
the interfaces can be extracted from the formula (\ref{U1super2}).

\subsubsection{Calculation of the Jacobian}
\label{anom2}

The Jacobian of the fermion path integral measure
for the $U(1)_V$ transformation (\ref{U1V})
in the odd dimensional case can be calculated in a similar
way as that for the even dimensional case in section \ref{anom}.
In particular, (\ref{Jacobian}), (\ref{sum}) and (\ref{cI}) can be
used for the $D=(2r+1)$ case with $\wh\cF$ and $\wh\cF'$ defined as
\begin{eqnarray}
\wh\cF=\frac{1}{2\Lambda^2}\gamma^\mu\gamma^\nu F_{\mu\nu}
+\frac{1}{\Lambda}\gamma^\mu D_\mu\wt m-\wt m^2\ ,
~~~
\wh\cF'=\frac{1}{2\Lambda^2}\gamma^\mu\gamma^\nu F_{\mu\nu}
-\frac{1}{\Lambda}\gamma^\mu D_\mu\wt m-\wt m^2\ .
\end{eqnarray}
Note that $\wh\cF'$ is obtained by replacing $\gamma^\mu$ with $-\gamma^\mu$
in $\wh\cF$. Therefore, when the matrix in the trace in (\ref{cI}) is expanded
with respect to $\gamma^\mu$ , only the terms with odd numbers of $\gamma^\mu$
can contribute. Furthermore, using the relation\footnote{
Here, $\gamma^{2r+1}$ is chosen to be the same as in (\ref{gamma5}).}
\begin{eqnarray}
\tr\left(
\gamma^{\mu_1}\cdots\gamma^{\mu_{2k+1}}
\right)
=\left\{
\begin{array}{ll}
~0&(k<r)\\
(2i)^r\epsilon^{\mu_1\cdots\mu_{2r+1}}&(k=r)
\end{array}
\right.
\ ,
\end{eqnarray}
we find that (\ref{cI2}) also holds for the odd dimensional
case, and
\begin{eqnarray}
\Tr_s\left(e^{\wh\cF}-e^{\wh\cF'}\right)d^{2r+1}\!x
=\Lambda^{-(2r+1)}(2i)^{r+1/2}\,\left[\Str\left(e^{\cF}\right)\right]_{2r+1}
+\cO(\Lambda^{-(2r+1)-1})\ ,
\end{eqnarray}
where $\cF$ is the superconnection of the odd type
given by (\ref{cF2}) with $T=\wt m=m/\Lambda$:
\begin{eqnarray}
\cF=\mat{F-\wt m^2,iD\wt m,iD\wt m,F-\wt m^2}\ .
\label{superF2}
\end{eqnarray}
Note that we have taken into account the $\sqrt{2}i^{-3/2}$ factor in
the definition of the supertrace `$\textrm{Str}$' for the odd case
(\ref{Str2}).
Then, we obtain
\begin{eqnarray}
\cI(x)d^{2r+1}\!x
=\left(\frac{i}{2\pi}\right)^{(2r+1)/2}
\left[\Str\left(e^{\cF}\right)\right]_{2r+1}
=\left[\ch(\cF)\right]_{2r+1}
\ .
\end{eqnarray}
This implies
\begin{eqnarray}
\log\cJ= -i\int \alpha(x)[\ch(\cF)]_{2r+1}\ ,
\label{logJodd}
\end{eqnarray}
which takes the same form as (\ref{U1super2})
for $D=2r+1$.

In particular, when the mass is a scalar matrix given by
\begin{eqnarray}
m=\mu(x) 1_N\ ,
\label{scalarm2}
\end{eqnarray}
with a real function $\mu(x)$, we have
\begin{eqnarray}
\ch(\cF)
=\frac{1}{\sqrt{\pi}}\,d\wt\mu\,
e^{-\wt\mu^2}\ch(F)\ ,
\end{eqnarray}
and
\begin{eqnarray}
\log\cJ
=-\frac{i}{\sqrt{\pi}}
\int d\wt\mu\,e^{-\wt\mu^2} \alpha(x)
\left[\ch(F)\right]_{2r}
\ ,
\label{logJscalar2}
\end{eqnarray}
where $\wt\mu\equiv \mu/\Lambda$.

\section{Applications}
\label{app}

\subsection{Anomalies on interfaces}
\label{interface}

In this section, we consider mass profiles with isolated zero loci,
which we call interfaces, and show
that the anomaly carried by the fermions localized
on the interfaces can be easily extracted by the formulas
obtained in section \ref{Derivation}.
As pointed out in \cite{Cordova:2019jnf,Hsin:2020cgg},
the anomaly of the localized modes implies the existence of
a diabolical point in the space of parameters of the theory,
which will be mentioned at the end of section \ref{highercodim}.

\subsubsection{Kink (codimension 1 interface)}
\label{kink}

We consider a $D=(2r+1)$-dimensional system
given by (\ref{Sodd}) with a kink-like mass profile as
\begin{eqnarray}
 m=\mu(y)1_N=uy1_N\ ,
\label{muy}
\end{eqnarray}
where $y\equiv x^{2r+1}$ is one of the spatial coordinates
and $u$ is a real parameter.
Since the mass $m$ diverges at $|y|\ra\infty$,
the operators $\cD^\dag\cD$ and $\cD\cD^\dag$ have discrete spectra
as required in section \ref{anom}.

To simplify the discussion, we assume that
the gauge field as well as $\alpha(x)$ are independent of $y$.
Then, the integration over $y$ in (\ref{logJscalar2}) can be done
and we obtain
\begin{eqnarray}
\log\cJ
=-i\,\sgn(u)
\int \alpha(x)\left[\ch(F)\right]_{2r}
\ ,
\label{logJdefect2}
\end{eqnarray}
where $\sgn(u)=u/|u|$ is a sign function and
the integration is taken over the $2r$-dimensional
space along $x^{1\sim 2r}$ directions.
Note that this result is independent of the cut-off $\Lambda$,
and hence it survives in the $\Lambda\ra\infty$ limit.
The dependence on the parameter $u$ is only through its sign.
Knowing this fact, for some purposes, it may be convenient to take
the $|u|\ra\infty$ limit as
\begin{eqnarray}
\lim_{|u|\ra\infty}\ch(\cF)
=\sgn(u)\delta(y)dy\,\ch(F)\ .
\end{eqnarray}
In fact, (\ref{logJdefect2}) does not depend
on the detail of the profile (\ref{muy}). As it is clear from
(\ref{logJscalar2}), we get the same result (\ref{logJdefect2})
for any function $\mu(y)$ satisfying
$\mu(y)\ra \pm \infty$ (or $\mu(y)\ra \mp \infty$)
as $y\ra\pm\infty$.

The expression (\ref{logJdefect2}) agrees with the anomaly for Weyl
fermions in a $2r$-dimensional spacetime. In fact, (\ref{logJdefect2})
is identical to (\ref{jacobian2}) with (\ref{U1anom}) provided we identify
$(F_+,F_-)=(F,0)$ for $u>0$ or $(F_+,F_-)=(0,F)$ for $u<0$.
We interpret this as the anomaly contribution from the Weyl fermions
localized on the interface at $y=0$.
As a check, it is easy to show
that there exist positive or negative chirality Weyl fermions
at the interface as the zero modes of the operator $\cD=\Dslash+m$
with $u>0$ or $u<0$, respectively.\cite{Jackiw:1975fn}
To see this, let us consider the Dirac equation $\cD\psi =0$,
where $\cD$ is defined in (\ref{Dslashodd}).
Working in the $A_{2r+1}=0$ gauge, this equation can be written as
\begin{eqnarray}
\Dslash^{(2r)}\psi+\gamma^{2r+1}\del_y\psi
+\mu(y)\psi=0\ ,
\label{zeromode}
\end{eqnarray}
where
\begin{eqnarray}
\Dslash^{(2r)}= \sum_{\mu=1}^{2r}
 \gamma^\mu(\del_\mu+A_\mu)
\label{D2r}
\end{eqnarray}
is the Dirac operator in the $2r$-dimensional space.
Then, we find a solution localized around $y=0$:
\begin{eqnarray}
\psi(\vec x,y)=e^{-\half |u|y^2}\psi^{(2r)}(\vec x)
\end{eqnarray}
where $\vec x=(x^1,\cdots,x^{2r})$ and $\psi^{(2r)}(\vec x)$
is the $2r$-dimensional Weyl fermion at the interface satisfying
\begin{eqnarray}
 \Dslash^{(2r)}\psi^{(2r)}=0\ ,~~~
\gamma^{2r+1}\psi^{(2r)}=\sgn(u) \psi^{(2r)}\ .
\label{normalizable}
\end{eqnarray}

Note, however, that the anomaly contribution of the localized Weyl
fermions are known to be canceled by the contribution from the bulk
via the anomaly inflow mechanism \cite{Callan:1984sa}.
Outside the region with $\mu(x)=0$, the one loop effective action
contains a term with the CS $(2r+1)$-form, whose
gauge variation precisely cancels the anomaly of the
localized fermions. Our result (\ref{logJdefect2}) can be
interpreted in two ways. One is that
the variation of CS-term simply vanishes
when the gauge field and the gauge variation are independent of $y$,
and (\ref{logJdefect2}) is the contribution of the localized fermion.
The other is that the anomaly of the localized fermion at $y=0$ is
canceled by the contribution from the CS-term,
but the variation of the CS-term also produces
the same amount of anomaly at $y=\pm\infty$, which gives
(\ref{logJdefect2}). We will make more comments on the relation
to the anomaly inflow below.

Let us next discuss the anomaly $(D+2)$-form (\ref{superID}).
Inserting (\ref{muy}) into (\ref{superID}), we obtain
\begin{eqnarray}
I_{2r+3}(A,\wt m)
=-2\sqrt{\pi} i\,
e^{-\wt m^2}d\wt m\,
[\ch(F)]_{2r+2}
=df(\wt m)
\,I_{2r+2}(A)\ ,
\label{I2r+3}
\end{eqnarray}
where $I_{2r+2}(A)\equiv -2\pi i[\ch(F)]_{2r+2}$ and
\begin{eqnarray}
f(x)\equiv\frac{1}{\sqrt{\pi}}\int_{0}^{x}e^{-y^2}dy
=\half\textrm{erf}(x)\ .
\label{fx}
\end{eqnarray}

A possible choice of $I_{2r+2}^0$ satisfying
the relation $I_{D+2}=dI_{D+1}^0$ in (\ref{descentD})
with $D=2r+1$ is
\begin{eqnarray}
I_{2r+2}^0(A,\wt m)=f(\wt m)I_{2r+2}(A)\ .
\label{I02r+2}
\end{eqnarray}
Since this is invariant under the $U(N)$ transformation,
we have $\delta_v I_{2r+2}^0(A,\wt m)=0$ and the anomaly
$I^1_{2r+1}$ related to $I_{2r+2}^0$ by the decent relation
(\ref{descentD}) vanishes.
However, this does not mean the $m$ dependent anomaly
$(D+2)$-form $I_{2r+3}(A,\wt m)$ is useless. In fact, we can extract
the information of the anomaly from the fermions localized at the
interface from (\ref{I2r+3}) as follows.

The point is that the factor $f(\wt m)$ in (\ref{I02r+2})
does not vanish but approaches $\pm \half\sgn(u)$ at $y=\pm\infty$.
Therefore, the relation $I_{2r+3}=dI^0_{2r+2}$ with a gauge invariant
$(2r+2)$-form $I^0_{2r+2}$ does not imply that $I_{2r+3}$ is trivial
as an element of cohomology with compact support.
To find the anomaly for the localized modes,
we decompose $I_{2r+2}^0$ in (\ref{I02r+2}) into
a local part that vanishes at $y\ra\pm\infty$ and a closed form
that does not contribute in the relation $I_{2r+3}=dI^0_{2r+2}$ as
\begin{eqnarray}
 I^0_{2r+2}(A,\wt m)=
I^{0\,\textrm{local}}_{2r+2}(A,\wt m)+d\omega_{2r+1}(A,\wt m)
\label{decomp}
\end{eqnarray}
with
\begin{eqnarray}
I^{0\,\textrm{local}}_{2r+2}(A,\wt m)
\equiv -df(\wt m) I^0_{2r+1}(A)\ ,~~~
\omega_{2r+1}(A,\wt m)
\equiv f(\wt m) I^0_{2r+1}(A)\ ,
\end{eqnarray}
where $I^0_{2r+1}(A)$ is the CS $(2r+1)$-form satisfying
$I_{2r+2}(A)=dI^0_{2r+1}(A)$.

We interpret $I_{2r+2}^{0\,\textrm{local}}(A,\wt m)$ as the part
that gives the anomaly localized at the interface.
Integrating $I_{2r+2}^{0\,\textrm{local}}(A,\wt m)$
over the $y$ direction, one obtains
a CS $(2r+1)$-form
\begin{eqnarray}
I_{2r+1}^{0\,\textrm{local}}(A)\equiv
-\int_{\{y\}} I_{2r+2}^{0\,\textrm{local}}(A,\wt m)
=\sgn(u)I^0_{2r+1}(A)\ ,
\end{eqnarray}
which is related to the anomaly $I_{2r}^{1\,\textrm{local}}(v,A)$
for the Weyl fermions localized at the interface by
the descent relation
$\delta_v I^{0\,\textrm{local}}_{2r+1}(A)=dI_{2r}^{1\,\textrm{local}}(v,A)$
in (\ref{descentD}).
Here, $\int_{\{y\}}$ denotes the integral over $y$.
The anomaly $(2r+2)$-form for the localized fermions
is given by
\begin{eqnarray}
I^\textrm{local}_{2r+2}(A)\equiv \int_{\{y\}}I_{2r+3}(A,\wt m)
=\sgn(u)I_{2r+2}(A)\ .
\label{Ilocal}
\end{eqnarray}

The second term in (\ref{decomp}) corresponds to the anomaly
contribution from the bulk that cancels the anomaly localized
at the interface around $y=0$ through the anomaly inflow \cite{Callan:1984sa}.
To see this explicitly, it is convenient to take the $|u|\ra\infty$ limit,
in which $f(\wt m)$ and $df(\wt m)$ approach a step function
and a delta function 1-form with support at $y=0$, respectively:
\begin{eqnarray}
 f(\wt m)\ra \half \sgn(u)\,\sgn(y)\ ,~~~
 df(\wt m)\ra \sgn(u)\delta(y)dy\ .
\end{eqnarray}
Then, $I_{2r+2}^{0\,\textrm{local}}(A,\wt m)$ is completely localized at $y=0$
and $\omega_{2r+1}$ becomes
\begin{eqnarray}
 \omega_{2r+1}(A,\wt m)\ra -\half\sgn(\wt m) I_{2r+1}^0(A)\ ,
\end{eqnarray}
which can be interpreted as the CS-term in the bulk
induced from the path integral of the massive fermions,
which precisely cancels the anomaly localized at the interface.

\subsubsection{Vortex (codimension 2 interface)}
\label{vortex}

Next, consider a $D=(2r+2)$-dimensional system (\ref{action})
with a vortex-type mass profile given by
\begin{eqnarray}
 m=\mu(z)1_N=uz1_N\ ,
\label{vortexmass}
\end{eqnarray}
where $z=x^{2r+1}-ix^{2r+2}$ and $u$ is a complex parameter.
Here, we assume that the gauge fields as well as
the parameter $\alpha$ are independent
of $z$, and satisfy $A_+=A_-\equiv A$ and $A_{2r+1}=A_{2r+2}=0$,
for simplicity.

Then, (\ref{logJscalar1}) implies
\begin{eqnarray}
\log\cJ
=-i\int  \alpha(x)
\left[\ch(F)\right]_{2r}\ .
\end{eqnarray}
This agrees with the anomaly of a $2r$-dimensional system with Weyl
fermions and it is interpreted as the anomaly contribution from
the Weyl fermion localized on the interface at $z=\ol z=0$.

Again, we can explicitly find localized Weyl fermions as
follows.\cite{Jackiw:1981ee,Witten:1984eb,Callan:1984sa}
For this purpose, it is convenient to choose
$\sigma^\mu=\gamma_{(2r)}^\mu$ ($\mu=1,\cdots,2r+1$)
and $\sigma^{2r+2}=-i1_{2^{r}}$, where $\gamma^\mu_{(2r)}$
($\mu=1,\cdots,2r$) are gamma matrices in $2r$-dimensions and
$\gamma_{(2r)}^{2r+1}$ is the chirality operator for them.
In this case, the Dirac equation $\cD\psi=0$ can be written as
\begin{eqnarray}
\Dslash^{(2r)}\psi_++2(P_+\del_{z}-P_-\del_{\ol z})\psi_+
+\ol u\ol z\psi_-&=&0\ ,\\
\Dslash^{(2r)}\psi_-+2(P_+\del_{\ol z}-P_-\del_{z})\psi_-
+u z\psi_+&=&0\ ,
\label{zeromode2}
\end{eqnarray}
where $\Dslash^{(2r)}$ is defined in (\ref{D2r}) and
$P_\pm\equiv \half(1_{2^r}\pm \gamma_{(2r)}^{2r+1})$
is a projection operator that projects to
positive/negative chirality spinors in $2r$-dimensions.
Then, we find a solution localized around $z=0$:
\begin{eqnarray}
 \psi_+(\vec x, z,\ol z)=\psi_-(\vec x, z,\ol z)=
e^{-\half u|z|^2}\psi^{(2r)}(\vec x)\ ,
\end{eqnarray}
where we have assumed $u$ to be real and positive without loss of
generality, and $\psi^{(2r)}$ is a positive chirality massless
Weyl fermion in $2r$-dimensions.\footnote{A negative chirality mode
is obtained when the mass is $m= u\ol z\, 1_N$,
which represents an anti-vortex.}

The role of the anomaly $(D+2)$-form (\ref{superI}) can be discussed
in a similar way as the codimension 1 interface considered
in section \ref{kink}.
For the mass profile (\ref{vortexmass}),
the anomaly $(D+2)$-form (with $D=2r+2$) becomes
\begin{eqnarray}
I_{2r+4}(A,\wt m)
=df_1(\wt m)\,I_{2r+2}(A)\ ,
\end{eqnarray}
where $I_{2r+2}(A)\equiv -2\pi i[\ch(F)]_{2r+2}$ is the anomaly
polynomial for a Weyl fermion in $2r$-dimensions and
 $f_1$ is a 1-form given by
\begin{eqnarray}
f_1(\wt m)
\equiv \frac{i}{4\pi}\left(1-e^{-|\wt m|^2}\right)
\left(d\log\wt m-d\log \wt m^\dag \right)\ .
\end{eqnarray}
Note that $f_1$ is non-vanishing at $|z|\ra\infty$, while
its derivative
\begin{eqnarray}
df_1(\wt m)= \frac{i}{2\pi}
d\wt m^\dag d\wt m\, e^{-|\wt m|^2}
\end{eqnarray}
decays exponentially as $|z|\ra\infty$, and
approaches a delta function 2-form with support at
$z=\ol z=0$ in the $u\ra\infty$ limit.
The integral of $df_1$ over the $z$-plane is normalized as
\begin{eqnarray}
\int df_1=1\ ,
\end{eqnarray}

The CS-form $I_{2r+3}^0(A,\wt m)$ satisfying
$I_{2r+4}(A,\wt m)=dI_{2r+3}^0(A,\wt m)$ can be chosen as
\begin{eqnarray}
I_{2r+3}^0(A,\wt m)=f_1(\wt m)I_{2r+2}(A)
=I_{2r+3}^{0\,\textrm{local}}(A,\wt m)+d\omega_{2r+2}(A,\wt m)
\end{eqnarray}
where
\begin{eqnarray}
I_{2r+3}^{0\,\textrm{local}}(A,\wt m)\equiv df_1(\wt m)I_{2r+1}^0(A)\ ,~~~
\omega_{2r+2}(A,\wt m)\equiv -f_1(\wt m)I_{2r+1}^0(A)\ .
\end{eqnarray}
Here, $I_{2r+1}^0(A)$ is the CS-form satisfying
$I_{2r+2}(A)=dI_{2r+1}^0(A)$.

The anomaly contribution of the fermions localized at the interface,
denoted as $I^{1\,\textrm{local}}_{2r}(A)$,
is related to
\begin{eqnarray}
I_{2r+1}^{0\,\textrm{local}}(A)\equiv
\int_{\{z,\ol z\}} I^{0\,\textrm{local}}_{2r+3}(A,\wt m)=
I^{0}_{2r+1}(A)\ .
\end{eqnarray}
where $\int_{\{z,\ol z\}}$ denotes the integral over
the $z$-plane,
by the descent relation $dI_{2r}^{1\,\textrm{local}}=
\delta_v I_{2r+1}^{0\,\textrm{local}}$.
In other words, it is characterized by the anomaly polynomial
\begin{eqnarray}
I_{2r+2}^\textrm{local}(A)\equiv
\int_{\{z,\ol z\}} I_{2r+4}(A,\wt m)=I_{2r+2}(A)\ .
\label{Ilocal2}
\end{eqnarray}
On the other hand,
$\omega_{2r+2}(A,\wt m)$ gives the bulk contribution of the anomaly
that cancels the anomaly on the interface.

\subsubsection{Interfaces of higher codimension}
\label{highercodim}

The discussion in sections \ref{kink} and \ref{vortex} can be
generalized to the cases with interfaces of higher codimensions.
We are interested in the interfaces with Weyl fermions on them.

A codimension $n$ interface in $D=(2r+n)$-dimensional spacetime
can be constructed by giving a mass of the form
\begin{eqnarray}
 m(x)= u\sum_{I=1}^{n}\Gamma^I x^I\ ,
\label{mx}
\end{eqnarray}
where $\Gamma^I$ ($I=1,2,\cdots,n$) are matrices
of size $N=2^{[(n-1)/2]}$ related to
$n$-dimensional gamma matrices $\wh\gamma^I$ by
\begin{eqnarray}
 \wh\gamma^I=\Gamma^I~~~(\mbox{for odd}~n~\mbox{and}~D)\ ,~~~
 \wh\gamma^I=\mat{~,\Gamma^I,\Gamma^{I\dag},~}~~~(\mbox{for even}~n~\mbox{and}~D)\ .
\end{eqnarray}
In this case, it can be shown that there is a Weyl fermion
on the interface at $x^1=\cdots=x^n=0$ obtained as a localized
fermion zero mode, as we have seen this explicitly in sections
\ref{kink} and \ref{vortex} for $n=1,2$. We will give an indirect
argument for this fact for general $n$
in connection to index theorems in section \ref{Callias}
and string theory interpretation in section \ref{string}.

It is also possible to get $k$ Weyl fermions by
replacing $\Gamma^I$ in (\ref{mx}) by $1_k\otimes\Gamma^I$ as
\begin{eqnarray}
 m(x)= u\sum_{I=1}^{n}1_k\otimes\Gamma^I x^I\ .
\label{mx2}
\end{eqnarray}
In this case, the gauge group is $U(kN)$ or
$U(kN)_+\times U(kN)_-$ for odd or even $D$, respectively,
and the vector-like $U(k)$ subgroup
of the form $g\otimes 1_N$ with $g\in U(k)$
is unbroken. Then, $k$ Weyl fermions coupled with $U(k)$
gauge field $a$ can be obtained by setting $U(kN)$ gauge field
$A$ as
\begin{eqnarray}
A=a\otimes 1_N\ .
\label{Ukgauge}
\end{eqnarray}
It is straightforward to check that
the anomaly for these Weyl fermions on the interface can be
obtained by inserting the mass profile (\ref{mx2})
and the gauge field (\ref{Ukgauge}) into our formulas
(\ref{superI})--(\ref{U1super}) and (\ref{superID})--(\ref{U1superD}).
In particular, the expressions (\ref{Ilocal}) and (\ref{Ilocal2})
of the anomaly $(2r+2)$-form for the localized fermions
are generalized as
\begin{eqnarray}
I_{2r+2}^\textrm{local}(a)\equiv
\int_{n} I_{2r+n+2}(A,\wt m)\ ,
\label{Ilocal3}
\end{eqnarray}
where $\int_n$ denotes the integral over $x^I$ ($I=1,2,\cdots,n$).
This agrees with the anomaly polynomial for $2r$-dimensional Weyl fermions
coupled to the $U(k)$ gauge field $a$.

As discussed in
\cite{Cordova:2019jnf,Hsin:2020cgg},
the anomaly contributions from fermion zero modes
localized  on the interfaces implies that there is at least
one point in the space of parameters of the theory, called
a diabolical point, at which the theory is not trivially gapped.
In our examples, it is of course clear that the massless point
$m=0$ is the diabolical point. However, since the anomaly
takes a discrete value, the existence of the diabolical point
is robust against continuous deformations of the theory.
In fact, as we have seen, the anomaly depends only on the asymptotic
behavior of the mass profile. The existence of the diabolical point
can be shown without examining the theory at the massless point.
This point is more explicit in the Callias-type index theorem
(\ref{indg}) discussed in section \ref{Callias}.

\subsection{Anomaly in spacetime with boundaries}
\label{bdry}

Since the fermions cannot propagate in a region with infinite mass,
it is possible to realize a spacetime with boundaries by considering
a spacetime dependent mass that blows up in some regions.
In this subsection, we discuss the anomaly driven by the boundary
condition imposed on the fermions, using our formulas
obtained in section \ref{Derivation}.

\subsubsection{Odd dimensional cases}
\label{oddbdry}

Let us first consider a $D=(2r+1)$-dimensional system of $N$ Dirac fermions
with $y\equiv x^{2r+1}$ dependent mass given by
\begin{eqnarray}
m(y)=\mu(y)1_N=\left\{
\begin{array}{cl}
(m_0+u'(y-L))1_N &(L<y)\\
m_01_N&(0\le y\le L)\\
(m_0+uy)1_N & (y< 0)
\end{array}
\right.\ ,
\label{massbdry}
\end{eqnarray}
where $u$, $u'$ and $m_0$ are real parameters.\footnote{
Strictly speaking, since $\del_y^2m$ has delta function singularities
at $y=0,L$, the assumption that we made above (\ref{cI2}) is not
satisfied. However, it can be shown that these singularities
do not contribute and the result is unchanged. Alternatively,
one could replace $\mu(y)$ with a smooth function with the same
asymptotic behavior as (\ref{massbdry}), which also gives
the same result.}
We assume that the gauge field is independent of $y$ in the
$y<0$ and $L<y$ regions,

When $|u|$ and $|u'|$ are large enough,
this system can be regarded as that of $N$ Dirac fermions with mass $m_0$
living in an interval $0\le y\le L$ with boundaries at $y=0$ and $y=L$.
The boundary conditions for the fermion fields follow from the requirement
that they do not blow up at $y\ra\pm\infty$.
The discussion around (\ref{zeromode})--(\ref{normalizable}) implies
that the corresponding boundary conditions are
\begin{eqnarray}
\left(\gamma^{2r+1} \psi-\sgn(u)\psi\right)|_{y=0}=0\ ,~~~
\left(\gamma^{2r+1} \psi-\sgn(u')\psi\right)|_{y=L}=0\ ,
\label{bc1}
\end{eqnarray}
which are equivalent to one of the boundary conditions considered
in \cite{Witten:2019bou}.

In this setup, the formula (\ref{logJscalar2}) implies that
the Jacobian is
\begin{eqnarray}
\log\cJ
=i\kappa_-
\int_{y=0} \alpha\left[\ch(F)\right]_{2r}
+i\kappa_+
\int_{y=L} \alpha\left[\ch(F)\right]_{2r}
\ ,
\label{logJbdry}
\end{eqnarray}
with
\begin{eqnarray}
\kappa_-=
\half\sgn(u)+f(\wt m_0)
\ ,~~~
\kappa_+=
\half\sgn(u')-f(\wt m_0)
\label{kappa}
\end{eqnarray}
where $\wt m_0\equiv m_0/\Lambda$ and $f(z)$ is the function defined
in (\ref{fx}), and $\alpha$ is assumed to be independent of $y$
in the $y<0$ and $L<y$ regions.
When the cut-off $\Lambda$ is sent to infinity, while keeping
$m_0$ finite, $f(\wt m_0)$ simply vanishes and we get
\begin{eqnarray}
 \kappa_-=\half\sgn(u)\ ,~~~
 \kappa_+=\half\sgn(u')\ .
\label{kappa2}
\end{eqnarray}

Note that each term in (\ref{logJbdry}) with (\ref{kappa2}) is
proportional to the anomaly contribution from a Weyl fermion in
$2r$-dimensions. However, since the coefficients $\kappa_\pm$ are not
integers, it is not possible to interpret this result as the
contribution from the Weyl fermions localized
at the boundaries.  This is because the wave function
of the fermions are not completely localized at the boundary
in our setup, unless we take the $|\wt m_0|\ra\infty$ limit.
One way to understand (\ref{kappa2}) is to use
the anomaly inflow argument given in section \ref{kink}.
Namely, the anomaly contributions from the modes localized
at $y=0$ and/or $y=L$ are canceled by the bulk CS-terms, but the
gauge variation of the (half-level) CS-terms implies non-vanishing
surface terms at $y=\pm\infty$, which gives (\ref{logJbdry})
with (\ref{kappa2}) as $\alpha$ and $F$ are independent
of $y$ for $y<0$ and $L<y$.
On the other hand, one can argue that $\kappa_\pm$ can be shifted as
$\kappa_\pm\ra\kappa_\pm\pm\beta$ by adding a local counterterm of the form
\begin{eqnarray}
 S_\textrm{c.t.}=\beta\int V[\ch(F)]_{2r}
\end{eqnarray}
where $V$ is the $U(1)$ gauge field, and including its gauge variation
in the Jacobian (\ref{logJbdry}). Therefore,
only the combination $\kappa_++\kappa_-=\half(\sgn(u)+\sgn(u'))$
is free from this ambiguity.

It is nonetheless useful to find the anomaly contribution
of the localized fermionic zero modes.
Assuming that $m_0$ is very large and
the $y$-dependence of the gauge field is negligible,
the solutions of the Dirac equation (\ref{zeromode})
in the region $0\le y<L$ are approximately a linear combination
of exponentially increasing and decreasing modes as
\begin{eqnarray}
 \psi(\vec x,y)\simeq e^{-m_0y}\psi_+^{(2r)}(\vec x)
+ e^{m_0y}\psi_-^{(2r)}(\vec x)\ ,
\end{eqnarray}
where $\psi_\pm^{(2r)}$ satisfies
\begin{eqnarray}
 \Dslash^{(2r)}\psi_\pm^{(2r)}=0\ ,~~~
\gamma^{2r+1}\psi_\pm^{(2r)}=\pm \psi_\pm^{(2r)}\ .
\end{eqnarray}
Then, the boundary conditions (\ref{bc1}) imply that there are
Weyl fermions localized near the boundary
with chirality $\sgn(u)$ and $\sgn(u')$ localized around $y=0$ and $y=L$,
if $\sgn(m_0)=\sgn(u)$ and $\sgn(m_0)=-\sgn(u')$, respectively.
The anomaly contributions of these localized modes are obtained by
formally taking the limit $|\wt m_0|\ra\infty$ in (\ref{kappa}),\footnote{
In this limit, only the localized zero modes are expected to contribute,
since the modes with energy greater than $\Lambda$
are suppressed by the heat kernel regularization (\ref{sum}).}
in which we have
\begin{eqnarray}
\kappa_-=\half(\sgn(u)+\sgn(m_0))\ ,~~~
\kappa_+=\half(\sgn(u')-\sgn(m_0))\ .
\label{kappa3}
\end{eqnarray}

\subsubsection{Even dimensional cases}
\label{evenbdry}

In this subsection, we consider a $D=2r$-dimensional
spacetime with boundaries realized by
the mass profile
\begin{eqnarray}
m(x)=\mu(y)g(x)=
\left\{
\begin{array}{cl}
u'(y-L)g(x) &(L<y)\\
0&(0\le y\le L)\\
uyg(x) & (y< 0)
\end{array}
\right.\ ,
\label{massbdry2}
\end{eqnarray}
where $y\equiv x^{2r}$, $g(x)\in U(N)$ and $u,u'\in\C$.
Since the phases of $u$ and $u'$ can be absorbed in $g(x)$,
we assume $u,u'>0$ without loss of generality.
We take a gauge with $A_{+y}=A_{-y}=0$ and
assume that the gauge fields $(A_+,A_-)$ and $g(x)$ are
independent of $y$ in the $y\le 0$ and $L\le y$ regions.
Since $\mu(y)$ vanishes in the region $0<y<L$, the $g(x)$ dependence
in this region drops out. Therefore, we can choose
$g(x)$ to be discontinuous in the region $\epsilon <y<L-\epsilon$
with $0<\epsilon\ll L$,
and the configuration of $g(x)$ at $y=0$ and $y=L$
can be topologically different.

As discussed around (\ref{bc1}) for the odd dimensional case,
by the requirement that the fermion fields do not blow up
at $y\ra\pm\infty$,
the boundary conditions corresponding to the mass profile
(\ref{massbdry2}) are obtained as
\begin{eqnarray}
\left(\gamma^{2r} \psi^g-\psi^g\right)|_{y=0,L}=0\ ,
\label{bc2}
\end{eqnarray}
where $\psi^g\equiv\left({g~~\atop~~1}\right)\psi
=\left(g\psi_+\atop\psi_-\right)$.\footnote{
This type of boundary condition with constant $g$
was introduced in the bag model of hadrons. \cite{Chodos:1974je}
The cases with $g=1$ or $g=i$ were considered recently
in \cite{Kurkov:2017cdz,Witten:2019bou}.}
Therefore, this system can be regarded as that of
massless $N$ Dirac fermions on the interval $0\le y\le L$ with
a boundary condition (\ref{bc2}).
Note that this boundary condition (\ref{bc2}) depends on the
spacetime coordinates through $g(x)$.
With fixed $g(x)$, the boundary condition (\ref{bc2})
breaks the $U(N)_+\times U(N)_-$ gauge symmetry
down to the $U(N)$ subgroup that consists of elements
$(U_+,U_-)\in U(N)_+\times U(N)_-$ with $U_-=gU_+g^{-1}$.
However, as it is evident
from our construction, the boundary condition (\ref{bc2})
is invariant under the gauge transformation
\begin{eqnarray}
 A_+\ra A_+^{U_+}\ ,~~~ A_-\ra A_-^{U_-}\ ,~~~
g\ra U_-\, g\, U_+^{-1} \ ,
\label{gauge}
\end{eqnarray}
and it makes sense to consider the anomaly with respect to
$U(N)_+\times U(N)_-$ even at the boundaries.

For this configuration, the field strength of the superconnection
(\ref{superF}) becomes
\begin{eqnarray}
\cF&=&
\mat{g^{-1},,,1_N}
\mat{F_+^g-\wt\mu^21_N,i(d\wt\mu 1_N-(A_--A_+^g)\wt\mu),
i(d\wt\mu 1_N+(A_--A_+^g)\wt\mu),F_--\wt\mu^21_N}
\mat{g,,,1_N}
\nn\\
&=&
\mat{g^{-1},,,1_N}\left(
-\wt\mu^21_{2N}+
F_+^ge^++F_-e^-
+id\wt\mu\sigma_1+\wt\mu(A_--A_+^g)\sigma_2
\right)
\mat{g,,,1_N}\ ,
\nn\\
\label{cFbdry}
\end{eqnarray}
where $\wt\mu\equiv \mu/\Lambda$, $A_+^g\equiv gA_+g^{-1}+gdg^{-1}$
and $F_+^g\equiv gF_+g^{-1}$.
The second line of (\ref{cFbdry}) is written
in the notation introduced in (\ref{cA}) with
$\sigma_1=\left(0\,1\atop1\,0\right)$
and $\sigma_2=\left(0\,-i\atop i~~0\right)$.
Then, we obtain
\begin{eqnarray}
\Str(e^{\cF})=e^{-\wt\mu^2}\Str\left(
e^{F_+^ge^++F_-e^-
+\wt\mu(A_--A_+^g)\sigma_2
}\left(1+id\wt\mu\sigma_1\right)
\right)\ ,
\end{eqnarray}
and, hence, the Jacobian (\ref{U1super2}) becomes
\begin{eqnarray}
\log\cJ=
-i\int_{0< y< L}\alpha\left[\ch(F_+)-\ch(F_-)\right]_{2r}
-i\int_{y=L}\alpha[\omega]_{2r-1}
+i\int_{y=0}\alpha[\omega]_{2r-1}\ ,
\label{logJbdry2}
\end{eqnarray}
where we have assumed that $\alpha$ is independent of $y$ in the
$y<0$ and $L<y$ regions, and defined
\begin{eqnarray}
\omega\equiv i\sum_{r\ge 1}
\left(\frac{i}{2\pi}\right)^r\int_0^\infty dt\, e^{-t^2}
\left[\Str\left(e^{F_+^ge^++F_-e^-+t(A_--A_+^g)\sigma_2}\sigma_1
\right)\right]_{2r-1}\ .
\label{omega}
\end{eqnarray}
This $\omega$ is a formal sum of differential forms
on the boundaries (\textit{i.e.} $y=0$ and $y=L$ planes).
The 1-form and 3-form components of $\omega$ are
\begin{eqnarray}
[\omega]_1&=&\frac{i}{2\pi}\Tr\left(A_--A_+^g\right)\ ,
\\
{[}\omega{]}_3&=&-\frac{1}{8\pi^2}
\Tr\left((A_--A_+^g)(F_-+F_+^g)
-\frac{1}{3}(A_--A_+^g)^3\right)
\ .
\end{eqnarray}
One can show that this a generalization of CS-forms satisfying
\begin{eqnarray}
d\omega|_{y=0,L}=\left(\ch(F_-)-\ch(F_+)\right)|_{y=0,L}\ ,
\label{domega}
\end{eqnarray}
and it is manifestly invariant under the gauge transformation
(\ref{gauge}).
To show (\ref{domega}), consider the $L\le y$ region and
note that $\omega$ at $y=L$ can be written as
\begin{eqnarray}
\omega|_{y=L}
=\int_{\{L\le y\}} \ch(e^\cF)\ ,
\label{omegachF}
\end{eqnarray}
where $\int_{\{L\le y\}}$ denotes the integration over $y$ with $L\le y$.
Then, applying the exterior derivative $d=d_x+d_y$,
where $d_x\equiv\sum_{\mu=1}^{2r-1}dx^\mu\del_\mu$
 and $d_y\equiv dy\,\del_y$, and using the fact that
$\ch(e^\cF)$ is a closed form, we obtain
\begin{eqnarray}
d\omega|_{y=L}
=\int_{\{L\le y\}} d_x\ch(e^\cF)
=-\int_{\{L\le y\}} d_y\ch(e^\cF)
=-\ch(e^\cF)|_{y=L}\ ,
\end{eqnarray}
which implies (\ref{domega}).

An important observation is that even if the gauge fields
are set to zero, (\ref{omega}) can be non-vanishing.
In fact, for $A_+=A_-=0$, we obtain
\begin{eqnarray}
[\omega]_{2r-1}=\left(\frac{-i}{2\pi}\right)^r\frac{(r-1)!}{(2r-1)!}
\Tr((gdg^{-1})^{2r-1})\ .
\label{omega0}
\end{eqnarray}
When the spacetime is of the form $S^{2r-1}\times\{y\}$,
the integral of this form over $S^{2r-1}$ gives a winding number
in $\pi_{2r-1}(U(N))$ represented by the map $g:S^{2r-1}\ra U(N)$.
If the winding number at $y=0$ and $y=L$ are the same,
a function $g:S^{2r-1}\times\{y\}\ra U(N)$ that interpolates
the configuration of $g$ at $y=0$ and $y=L$ can be found and
the Jacobian (\ref{logJbdry2}) can be canceled by the gauge
variation of a local counterterm
\begin{eqnarray}
S_\textrm{c.t.}=-\int_{0<y<L}V [\omega]_{2r-1}\ ,
\end{eqnarray}
where $V$ is the $U(1)_V$ gauge field and $[\omega]_{2r-1}$
is given by (\ref{omega0}).
However, when the winding numbers at $y=0$ and $y=L$ are different,
this is not allowed and there is an anomaly.

Another interesting situation is the case with $g(x)=e^{i\phi(x)}1_N$
and $A\equiv A_+=A_-$. In this case, the formula (\ref{omega}) implies
\begin{eqnarray}
\omega=-\frac{d\phi}{2\pi} \ch(F)\ .
\end{eqnarray}
Therefore, when the spacetime is of the form
$S^1\times S^{2r-2}\times\{y\}$ and
the winding number of $e^{i\phi}$ on $S^1$
for $y=0$ and $y=L$ are different,
there is an anomaly for the $U(1)_V$ symmetry in the presence of
a non-vanishing background vector-like gauge field on $S^{2r-2}$.

\subsection{Index theorems}
\label{appindex}

{}From (\ref{norm}) and the first expression in (\ref{sum}), we find
that the integral of $\cI(x)$ gives the index of operator $\cD$:
\begin{eqnarray}
\int d^{D}\!x\,\cI(x)=n_\varphi-n_\phi
=\dim\ker\cD-\dim\ker\cD^\dag\equiv
\Ind(\cD)\ ,
\end{eqnarray}
and the result (\ref{Iint}) implies an index theorem
written in terms of the superconnection:\footnote{
A quick way to get the expression of the index from the results
of the Jacobian in the previous sections is to set $\alpha=i$
in $\log\cJ$ as $\Ind(\cD)=\log\cJ|_{\alpha=i}$.
}
\begin{eqnarray}
\Ind(\cD)=\int [\ch(\cF)]_{D}\ .
\label{index}
\end{eqnarray}
When we set $m=0$ and $A_-=0$ in an even dimensional case,
this formula reduces to
a more familiar form of the Atiyah-Singer (AS) index theorem:
$\Ind(\Dslash)=\int\ch(F_+)$.
Thus, (\ref{index}) is a generalization of the AS index theorem,
which includes spacetime dependent mass and is supposed to hold
even when the spacetime manifold is odd dimensional
and/or non-compact, provided that
the spectra of $\cD\cD^\dag$ and $\cD^\dag \cD$ are discrete.

Here, we discuss some of the implications of this formula.
We will not try to make the statements mathematically
rigorous.\footnote{See, e.g., \cite{Kahle} for
mathematically rigorous description of index theorems
using the superconnection.}
Nevertheless, we hope they are useful and worth mentioning.

\subsubsection{Atiyah-Patodi-Singer index theorem}
\label{APStheorem}

The Atiyah-Patodi-Singer (APS) index theorem \cite{APS}
is an index theorem for a Dirac operator on an even dimensional
manifold $N$ with boundary, stated as
\begin{eqnarray}
\Ind(\Dslash)=\int\ch(F)\wh A(R)-\half\eta(i\Dslash_b)\ ,
\label{APS}
\end{eqnarray}
where $\Dslash$ is a Dirac operator on $N$,
$\eta(i\Dslash_b)$ is the eta invariant of a Dirac
operator on the boundary denoted as $\Dslash_b$ (see (\ref{eta})).\footnote{
See \cite{Fukaya:APS,Vassilevich:APS,Kobayashi:2021jbn}
for recent physicists-friendly formulations and derivations.
See also \cite{AlvarezGaume:1984nf} and Appendix \ref{app:APS}.
}

In this subsection, we first generalize (\ref{APS}) to include the
spacetime dependent mass $m$ and
then apply it to the system considered in section \ref{evenbdry}.
Let us consider a system in section \ref{even}
with $D=2r$-dimensional spacetime of the form $N=M\times I$,
where $M$ is a $(2r-1)$-dimensional manifold with coordinates $x^\mu$
($\mu=1,2,\cdots,2r-1$) and $I=[y_-,y_+]\subset\R$ is an interval
parameterized by $y\equiv x^{2r}\in I$.
For simplicity, as in the previous sections, we assume $M$ to be flat and
the $\wh A$-genus is omitted.

It is convenient to choose $\sigma^\mu$ in (\ref{cD})
such that $\sigma^{2r}=1_{2^{r-1}}$ and
$\sigma^{\mu}=i\gamma^\mu$ ($\mu=1,2,\cdots,2r-1$) with $\gamma^\mu$
being the $(2r-1)$-dimensional gamma matrices. Then the operator $\cD$
defined in (\ref{cD}) and its conjugate $\cD^\dag$ can be written as
\begin{eqnarray}
 \cD=\del_y+H_y\ ,~~~ \cD^\dag=-\del_y+H_y\ ,
\label{cDHy}
\end{eqnarray}
in the $A_{+y}=A_{-y}=0$ gauge, where
\begin{eqnarray}
 H_y\equiv\mat{-i\Dslash^{(2r-1)}_+,m^\dag,m,i\Dslash^{(2r-1)}_-}\ ,
\end{eqnarray}
\begin{eqnarray}
\Dslash^{(2r-1)}_+=\sum_{\mu=1}^{2r-1}\gamma^\mu(\del_\mu+A_{+\mu})\ ,~~~
\Dslash^{(2r-1)}_-=\sum_{\mu=1}^{2r-1}\gamma^\mu(\del_\mu+A_{-\mu})\ .
\end{eqnarray}
Note that although $H_y$ is $y$-dependent,
it does not contain the derivative with respect to $y$
and it can be regarded as a Hermitian operator acting on spinors on $M$.
Here, the mass $m$ can depend on both $x^\mu$ and $y$.
When $M$ is non-compact, the mass should diverge at infinity,
as the examples considered in sections \ref{interface} and \ref{bdry},
so that $H_y$ has a discrete spectrum.

The eta invariant of a Hermitian operator $H$ is defined as
\begin{eqnarray}
\eta(H)\equiv\lim_{s\ra 0}\eta(s,H)\ ,~~~
 \eta(s,H)\equiv\frac{2}{\Gamma((s+1)/2)}
\int_0^\infty dt\, t^s\,\Tr_{\cH}\left(H e^{-t^2H^2}\right)\ ,
\label{eta}
\end{eqnarray}
where the trace $\Tr_{\cH}$ is over the Hilbert space $\cH$ on which
the operator $H$ is acting and $s\ra 0$ limit is taken after
analytic continuation of $\eta(s,H)$ on the
complex $s$-plane. \cite{APS}
$\eta(s,H)$ can be written as a sum over eigenvalues $\lambda$ of $H$ as
\begin{eqnarray}
 \eta(s,H)=\sum_{\lambda}\sgn(\lambda)|\lambda|^{-s}\ .
\label{eta2}
\end{eqnarray}
Here and in the following, we assume that $H$ does not have a zero
eigenvalue, whenever it is used in $\eta(H)$ or $\eta(s,H)$.
For the massless case, the eta invariant of $H_y$ reduces to the difference
of the eta invariant of the Dirac operators $i\Dslash_+^{(2r-1)}$
and $i\Dslash_-^{(2r-1)}$ as
\begin{eqnarray}
\eta(H_y)|_{m=0}=-\eta(i\Dslash_+^{(2r-1)})+\eta(i\Dslash_-^{(2r-1)})
\label{etamassless}
\end{eqnarray}

Then, as it is explained in Appendix \ref{app:APS}, the index of $\cD$
is given by
\begin{eqnarray}
 \Ind(\cD|_I)=\lim_{\Lambda\ra\infty}\int_{y_-<y<y_+}[\ch(\cF)]_{2r}+\half\left[
\eta(H_{y})\right]^{y=y_+}_{y=y_-}\ ,
\label{APS2}
\end{eqnarray}
where $\cF$ is the field strength of the superconnection (\ref{superF})
with $\Lambda\ra\infty$ taken after the integration,
 $[f(y)]^{y=y_+}_{y=y_-}\equiv f(y_+)-f(y_-)$ and
$\Ind(\cD|_I)$ denotes the index of the operator $\cD$ acting on spinors
on $M\times I$ with the following APS boundary conditions.
For the operator $\cD$, when the wave function at $y=y_\pm$
is expanded with respect to eigenfunctions of $H_{y_\pm}$,
the components with the negative (for $y=y_+$) or positive
(for $y=y_-$) eigenvalues of $H_{y_\pm}$ have to vanish.
The conditions for the operator $\cD^\dag$ are the same as $\cD$
with the replacement $H_{y_\pm}\ra-H_{y_\pm}$.
These boundary conditions follow from the requirement that
wave function of the fermion does not blow up at $y\ra\pm\infty$,
when the system is extended to the $y<y_-$ and $y_+<y$ regions
with a $y$-independent configuration for $y\le y_-$ and $y_+\le y$.
(See Appendix \ref{app:APS}.)

Let us apply (\ref{APS2}) to the system considered in
section \ref{evenbdry}. Using (\ref{etamassless}),
the formula (\ref{APS2}) with $[y_-,y_+]=[0,L]$ becomes
\begin{eqnarray}
\Ind(\cD|_{[0,L]})
=\int_{0< y< L}\left[\ch(F_+)-\ch(F_-)\right]_{2r}
-\half\left[
\eta(i\Dslash_+^{(2r-1)})-\eta(i\Dslash_-^{(2r-1)})\right]^{y=L}_{y=0}\ ,
\label{indinterval1}
\end{eqnarray}
which is the APS index theorem for the massless Dirac operator
defined by $\cD|_{m=0}$ with the APS boundary conditions.
On the other hand, for $[y_-,y_+]=[-\infty,+\infty]$,
(\ref{index}) can be used, and from (\ref{logJbdry2}), we obtain
\begin{eqnarray}
\Ind(\cD)=
\int_{0< y< L}\left[\ch(F_+)-\ch(F_-)\right]_{2r}
+\int_{y=L}[\omega]_{2r-1}
-\int_{y=0}[\omega]_{2r-1}\ .
\label{indinterval2}
\end{eqnarray}
This is interpreted as the index theorem for the massless fermions
in the interval $[0,L]$ with the boundary condition given by (\ref{bc2}).

The difference between (\ref{indinterval1}) and (\ref{indinterval2})
can be evaluated by applying (\ref{APS2}) to the cases with
$[y_-,y_+]=[L,+\infty]$ and $[-\infty,0]$ (More precisely,
(\ref{indplus}) and (\ref{indminus}) with $\eta_0=0$.) :
\begin{eqnarray}
\Ind(\cD|_{[L,+\infty]})&=& \int_{y=L}[\omega]_{2r-1}
+\half\left(\eta(i\Dslash_+^{(2r-1)})-\eta(i\Dslash_-^{(2r-1)})\right)
\Big|_{y=L}\ ,
\nn\\
\Ind(\cD|_{[-\infty,0]})&=& -\int_{y=0}[\omega]_{2r-1}
-\half\left(\eta(i\Dslash_+^{(2r-1)})-\eta(i\Dslash_-^{(2r-1)})\right)
\Big|_{y=0}\ .
\end{eqnarray}
In particular, it implies a well-known relation between
eta invariant of a Dirac operator and the CS-form
$\omega$ defined by (\ref{omega}):
\begin{eqnarray}
\int [\omega]_{2r-1}=
\half\left(\eta(i\Dslash_-^{(2r-1)})-\eta(i\Dslash_+^{(2r-1)})\right)
~~~(\textrm{mod}~\Z)\ .
\end{eqnarray}

\subsubsection{Callias-type index theorem}
\label{Callias}

To illustrate the importance of the mass parameter
(or the Higgs field) in the formula (\ref{index}), let us consider
the case where the gauge fields are turned off.
The spacetime manifold is chosen to be a $D$-dimensional plane $\R^{D}$,
where $D$ can be either even or odd.
In order to have discrete spectrum, we assume that the mass diverges
at infinity. To be specific, the asymptotic behavior
of the mass is assumed to be as
\begin{eqnarray}
\wt m\ra r g(x)~~~(\mbox{as}~r\ra\infty)\ ,
\end{eqnarray}
where $r=\sqrt{x_\mu x^\mu}$ is the radial coordinate of $\R^D$
and $g(x)\in U(N)$ is a unitary matrix that only depends
on the angular coordinates of $\R^D$.
For odd $D$, $g(x)$ is also required to be
Hermitian.\footnote{Here, we assume $g(x)$ to be unitary
for computational simplicity. However, this condition can be relaxed
to $g(x)\in GL(N,\sC)$, as an invertible matrix
(or invertible Hermitian matrix)
can be continuously deformed to a unitary matrix
(or unitary Hermitian matrix, respectively),
keeping the invertibility.
}

Then, the right hand side of (\ref{index}) can be easily evaluated
by using (\ref{eF1-eF0_2}). The result is
\begin{eqnarray}
\Ind(\cD)= \int\ch(\cF)
=\left\{
\begin{array}{cc}
\displaystyle
\left(\frac{-i}{2\pi}\right)^{\frac{D}{2}}
\frac{\left(\frac{D}{2}-1\right)!}{(D-1)!}
\int_{S^{D-1}}
\Tr\left((gdg^{-1})^{D-1}\right)\ ,&(\mbox{for even $D$})
\vspace{2ex}
\\
\displaystyle
\left(\frac{i}{8\pi}\right)^{\frac{D-1}{2}}\frac{1}{2\left(\frac{D-1}{2}\right)!}
\int_{S^{D-1}}
\Tr\left((dg)^{D-1}g\right)\ ,&(\mbox{for odd $D$})
\end{array}
\right.\ ,
\nn\\
\label{indg}
\end{eqnarray}
where $S^{D-1}$ is the sphere at $r\ra\infty$.
The former (even $D$ case) is the same as the integral of (\ref{omega0})
over $S^{D-1}$ and the latter (odd $D$ case) agrees with expression
of the index for Callias's index theorem.\cite{Callias:1977kg}

We can apply these formulas for the configuration given by (\ref{mx}),
in which $g(x)$ is given by
\begin{eqnarray}
 g(x)=\frac{1}{r} \sum_{I=1}^{n}\Gamma^I x^I\ .
\end{eqnarray}
Inserting this into (\ref{indg}), we obtain $\Ind(\cD)=(-1)^{[\frac{D-1}{2}]}$,
which is consistent with the fact that there is
a fermionic zero mode as suggested in section \ref{highercodim} from
the existence of the anomaly.

\section{Relation to string theory}
\label{string}

Many of our results have natural interpretation in string theory.
In fact, it is well-known that the CS-terms
for unstable D-brane systems (D-brane - anti-D-brane systems
and non-BPS D-branes)
can be written by using superconnections\footnote{
As in the previous sections, we omit the terms with
curvature represented by the $\wh A$-genus.
}
\cite{Kennedy:1999nn,Kraus:2000nj,Takayanagi:2000rz,Alishahiha:2000du} as
\begin{eqnarray}
 S^\textrm{D9}_\textrm{CS}=\int C\, \ch(\cF)\ ,
\label{CS}
\end{eqnarray}
where $C$ is a formal sum of Ramond-Ramond (RR) $n$-form fields
($n$ is even or odd for type IIA or type IIB string theory, respectively.)
and $\cF$ is the field strength of the superconnection for the gauge
field and tachyon field on them,\footnote{See \cite{Hashimoto:2015iha}
for a generalization.}
and it is natural to anticipate the appearance of the superconnection
in anomaly analysis of quantum field theory counterparts.

An easy way to realize even dimensional systems having fermions
with manifest chiral symmetry is to consider a D$p$-brane
($p=-1,1,3,5,7$) with D9-branes and \AD9-branes
in type IIB string theory.\cite{Sugimoto:2004mh}\footnote{
A T-dual version ($N_c$ D4-branes with $N_f$ D8-\AD8 pairs)
is used in \cite{Sakai:2004cn} to realize QCD
in string theory.} On the D$p$-brane world-volume,
$(p+1)$-dimensional fermions are obtained in the spectrum
of $p$-9 strings and $p$-$\ol 9$ strings. Here, a $p$-$p'$ string
is an open string stretched between a D$p$-brane and a D$p'$-brane,
and $\ol p$ corresponds to a \AD{p}-brane.
It can be shown that $p$-9 strings and $p$-$\ol 9$ strings
create positive and negative chirality Weyl fermions, respectively.
When we have $N$ D9-\AD9 pairs,
there are $N$ flavors of fermions and the $U(N)\times U(N)$ gauge symmetry
associated with the D9-\AD9 pairs corresponds to the
$U(N)_+\times U(N)_-$ chiral symmetry for the $(p+1)$-dimensional
system realized on the D$p$-brane.
The CS-term of the D9-\AD9 system is written as (\ref{CS})
with $\cF$ being the field strength of the superconnection
of the even type (\ref{cF}), in which $A_+$ and $A_-$ are
the $U(N)\times U(N)$ gauge fields  given by 9-9 strings
and $\ol 9$-$\ol 9$ strings, respectively, and $T$ is the tachyon
field obtained by 9-$\ol 9$ strings. The tachyon field $T$ is in
the bifundamental representation of the $U(N)\times U(N)$ symmetry.
It couples with the fermions with Yukawa interaction and the value
of the tachyon field plays the role of the mass of the fermions.
When $|T|\ra\infty$, the fermions decouple, which correspond to
the annihilation of the D9-\AD9 pairs.

Similarly, odd dimensional systems with $N$ Dirac fermions can be obtained
by placing a D$p$-brane ($p=0,2,4,6,8$) with $N$ non BPS D9-branes
in type IIA string theory.
In this case, the CS-term for the non-BPS D9-branes is given by
(\ref{CS}), where $\cF$ is the odd type given by (\ref{cF2}).
Here, $A$ and $T$ in $\cF$ are the $U(N)$ gauge field
and the tachyon field, respectively, on the non-BPS D9-branes.
The tachyon field $T$ is a Hermitian matrix of size $N$ and transforms
as the adjoint representation of the $U(N)$ symmetry. There are
$N$ Dirac fermions in the spectrum of $p$-$9$ strings, which
are in the fundamental representation of $U(N)$, and
the value of the tachyon field corresponds to the mass
of the fermions.

Although the CS-term (\ref{CS}) for the unstable D-brane system
was originally derived by the computation of the interaction with the
RR fields, it can be determined by the requirement of
the anomaly cancellation as argued in
\cite{Green:1996dd,Cheung:1997az,Minasian:1997mm,Morales:1998ux,
Scrucca:1999uz,Szabo:2001yd}.
For the brane configuration above, the standard argument shows that
the anomaly contribution from the CS-term for the unstable
D9-branes (\ref{CS}) and the D$p$-brane
\begin{eqnarray}
 S_\textrm{CS}^{\textrm{D}p}=\int_{M} C\,\ch(f)\ ,
\end{eqnarray}
where $M$ is the $D=p+1$-dimensional D$p$-brane world-volume and
$\ch(f)=\exp\left(\frac{i}{2\pi}f\right)$ is the Chern character
for the $U(1)$ gauge field on it,
is given by the anomaly $(D+2)$-form of the form\footnote{
To be more precise, we should consider an anomaly 12-form
of the form
$2\pi i \left[\ch(\cF)\ch(f)\delta_{9-p}\right]_{12}$,
where $\delta_{9-p}$ is a delta function $(9-p)$-form supported on $M$.
}
\begin{eqnarray}
2\pi i \left[\ch(\cF)\ch(f)\right]_{D+2}\ .
\label{Dpanom}
\end{eqnarray}
Note that (\ref{Dpanom}) can be written as
$2\pi i\left[\ch(\cF)\right]_{D+2}$ by absorbing the $U(1)$ gauge
field on the D$p$-brane into the $U(1)_V$ part of the gauge field
of the unstable D9-brane system. This contribution is supposed
to cancel the anomaly contribution from the fermions,
which is indeed the case with  our proposal (\ref{superI}) and
(\ref{superID}), provided that the tachyon field is identified
with the mass as $T=\wt m$. From the dimensional analysis,
the cut-off $\Lambda$ is of the order of the string scale,
though the precise relation between $\Lambda$ and the string
length $l_s$ is not clear.

The argument above suggests that the anomaly is characterized
by the anomaly $(D+2)$-form written in terms of the Chern character
of the superconnection. However,
as discussed in section \ref{kink} and \ref{vortex},
since the $T$ dependent part of the anomaly $(D+2)$-form drops
out in the naive use of the anomaly descent relation,
it is important to have more evidence for this statement.
To this end, let us show that the analysis in section \ref{interface}
is consistent with the D-brane descent relation
\cite{Sen,Witten:1998cd,Horava:1998jy}.\footnote{
See, e.g., \cite{Sen:1999mg,Olsen:1999xx} for reviews.}

It is known that a D$q$-brane ($q$ is even/odd for type IIA/IIB)
localized at $x^I=0$ ($I=1,2,\cdots,9-q$) can be realized as a
soliton in the unstable D9-brane system by choosing the tachyon
field as in (\ref{mx}) with $n\equiv 9-q$ and $u\ra\infty$.
\cite{Witten:1998cd,Horava:1998jy}
In fact, the tachyon configuration with (\ref{mx}) is related to the
generator of K-groups $K(\R^{n})\simeq\Z$ or $K^1(\R^{n})\simeq\Z$
for even or odd $n$, respectively, given by the Atiyah-Bott-Shapiro
construction \cite{Atiyah:1964zz}, and these K-groups correspond
to the D$q$-brane charge.
When we have the D$p$-brane extended along $x^\mu=0$ ($\mu=0,1,\cdots,p$)
with $9-q\le p$, the D$q$-brane corresponds to the codimension
$(9-q)$ interface considered in section \ref{highercodim}.
($q=8$ and $q=7$ correspond to the kink and vortex considered
in sections \ref{kink} and \ref{vortex}, respectively.)

For this intersecting D$p$-D$q$ system, it can be shown that there
is a Weyl fermion localized at the $(p+q-8)$-dimensional intersection
in the spectrum of $p$-$q$ strings,  obtained by quantization
of the open string. This is consistent with the analysis of the
localized fermionic zero modes in section \ref{interface}.

Furthermore, we can obtain $k$ D$q$-branes with $U(k)$ gauge
field $a$ on them by choosing
the tachyon and gauge fields as (\ref{mx2}) and (\ref{Ukgauge}).
Then, one can show that the CS-term for the D$q$-brane is
reproduced from (\ref{CS}) by inserting
(\ref{mx2}) and (\ref{Ukgauge}) into (\ref{CS}) and integrating over
the transverse space \cite{Asakawa:2002ui}
(see also \cite{Olsen:1999xx}), which corresponds to the procedure
in (\ref{Ilocal3}).
As the anomaly contribution from the CS-terms for the
D$p$-brane and D$q$-branes precisely cancels that of
the Weyl fermions created by the $p$-$q$ strings,
the anomaly polynomial for these Weyl fermions
is given by (\ref{Ilocal3}), which is completely parallel
to the discussion in section \ref{interface} for the localized
fermionic zero modes.

\section{Conclusion}
\label{conclusion}

In this paper,
we have investigated the anomaly of fermions with spacetime dependent
mass. It was found in section \ref{Derivation} that the $U(1)_V$ anomaly and
the anomaly $(D+2)$-form are written with the Chern character
of the superconnection in both even and odd dimensional cases
as (\ref{superID}), (\ref{U1superD}), (\ref{superID}) and (\ref{U1superD}).
Applications of these formulas were discussed
in section \ref{app}. In section \ref{interface}, we considered the
interfaces made by the spacetime dependent mass
on which Weyl fermions are localized and confirmed that our formulas
can be used to extract the anomaly of these Weyl fermions.
The boundaries of spacetime realized by making the mass large in some
regions were studied in section \ref{bdry}.
A notable example was a system with the spacetime dependent boundary conditions
(\ref{bc2}) considered in section \ref{evenbdry}.
It was found that there are contributions to the anomaly from the
boundaries, even when the gauge fields are turned off.
Implications to the index theorems were discussed in section \ref{appindex},
in which the AS and APS index theorems for the operator $\cD$
defined in (\ref{cD}) and (\ref{Dslashodd}) were given,
and the application to the Callias-type index theorems
was briefly described. Finally, in section \ref{string},
we pointed out that the system of fermions with
spacetime dependent mass can be realized in string theory
and our formulas of anomaly are consistent with the anomaly
cancellation via the anomaly inflow from the CS-term
of the unstable D9-brane systems.

In this paper, we have considered complex Dirac fermions.
An obvious interesting problem would be to generalize our
discussion to systems with real or pseudo-real fermions,
for which there are 8 families of theories. For this purpose,
the concept of real superconnections and their realization on
unstalbe D-brane systems considered in \cite{Asakawa:2002ui} would
be useful.

Although we have seen that the formulas for the anomaly with the
superconnection are quite useful in some applications, we have
not explored much on the significance of the superalgebra acting on it.
It would be interesting if a deeper meaning behind this structure
could be uncovered.\footnote{
See, e.g., \cite{Schwimmer:1981yy,Thierry-Mieg:2020jqo} for
the works in this direction.}

\section*{Acknowledgement}
We would like to thank
S. Aoki, H. Fukaya, M. Honda, H. Shimada, Y. Tanizaki, S. Terashima
and K. Yonekura for useful discussions. We are especially grateful
to H. Fukaya and K. Yonekura for valuable comments on a draft of this paper.
We also appreciate useful discussions during YITP workshop,
``Topological Phase and Quantum Anomaly 2021'' (YITP-T-21-03).
In particular, we thank K.~Ohmori and M.~Yamashita for letting us know a relevant
paper \cite{Kahle} and helpful discussion on index theorems.
The work of SS was supported by JSPS KAKENHI (Grant-in-Aid for Scientific
Research (B)) grant number JP19H01897.
The work of HK was supported by the establishment of university
fellowships towards the creation of science technology innovation.

\appendix
\section{The APS index theorem}
\label{app:APS}

In this appendix, we give a heuristic derivation of (\ref{APS2})
following the argument given in the appendix of
\cite{AlvarezGaume:1984nf}.
The setup is the same as that of section \ref{APStheorem}.
As mentioned below (\ref{APS2}), we extend the system
to $-\infty<y<+\infty$ by choosing a $y$-independent configuration
in the regions $y\le y_-$ and $y_+\le y$.

First, we derive one of the key relations:
\begin{eqnarray}
\int_M d^{2r-1}x\, \cI(x)
=\frac{1}{\sqrt{\pi}}\lim_{\Lambda\ra\infty}
\frac{1}{\Lambda}
\Tr_{\cH}\left((\del_y H_y)\, e^{-\frac{1}{\Lambda^2}H_y^2}
\right)\ .
\label{key1}
\end{eqnarray}
Inserting
\begin{eqnarray}
\cD^\dag\cD=H_y^2-\del_y^2-\del_y H_y\ ,~~~
\cD\cD^\dag=H_y^2-\del_y^2+\del_y H_y\ ,
\end{eqnarray}
into (\ref{sum}), we obtain
\begin{eqnarray}
&&\int_M d^{2r-1}x\,\cI(x)
\nn\\
&=&
\lim_{\Lambda\ra\infty}\Lambda\int\frac{d\wt k}{2\pi}\,
e^{-\wt k^2}
\Tr_\cH\left(
e^{\frac{1}{\Lambda^2}\del_y^2+\frac{2i}{\Lambda}\wt
k\del_y-\frac{1}{\Lambda^2}(H_y^2-\del_yH_y)}
-e^{
\frac{1}{\Lambda^2}\del_y^2+\frac{2i}{\Lambda}\wt k\del_y
-\frac{1}{\Lambda^2}(H_y^2+\del_yH_y)}
\right)\ ,
\nn\\
\end{eqnarray}
where $\wt k=k_{2r}/\Lambda$.
As we did around (\ref{cI}),
we expand the right hand side with respect to $1/\Lambda$
regarding $\wt k$ and $H_y/\Lambda$ to be of $\cO(1)$.
The leading term in the $1/\Lambda$ expansion gives (\ref{key1}).

On the other hand, (\ref{eta}) implies
\begin{eqnarray}
\del_y\eta(H_y)
&=&\frac{2}{\sqrt{\pi}}
\int_0^\infty dt\,\Tr_\cH\left((\del_y H_y)(1-2t^2 H_y^2)\,
 e^{-t^2H_y^2}\right)
\nn\\
&=&\frac{2}{\sqrt{\pi}}
\int_0^\infty dt\,\del_t\Tr_\cH\left(t(\del_y H_y)\,e^{-t^2H_y^2}\right)
\nn\\
&=&-\frac{2}{\sqrt{\pi}}\lim_{\epsilon\ra 0}
\,\Tr_\cH\left(\epsilon(\del_y H_y)\,e^{-\epsilon^2H_y^2}\right)
\ .
\end{eqnarray}
Combining this with (\ref{key1}), we obtain
\begin{eqnarray}
-\half\del_y\eta(H_y)= \int_{M} d^{2r-1}x\,\cI(x)\ ,
\label{key2}
\end{eqnarray}
which can be used when $H_y$ does not have a zero eigenvalue.

Let us assume that $H_y$ has zero eigenvalues at
finite values of $y$ denoted as $y_i$ ($i=1,2,\cdots,k$) with
$y_-<y_1<y_2<\cdots<y_{k}<y_+$.
{}From the expression in (\ref{eta2}), we see that the value of
$\eta(H_y)$ jumps by $+2$ or $-2$ at $y=y_i$
when one of the eigenvalues of $H_y$ crosses zero from below
or above, respectively, while increasing $y$ from $y=y_i-\epsilon$
to $y=y_i+\epsilon$ with a positive small parameter $0<\epsilon\ll 1$.
It is known that the index of the operator $\cD$ is
equal to a half of the sum over these jumps \cite{APS}:
\footnote{
This fact can be easily understood in the adiabatic
limit,\cite{Callan:1977gz,Kiskis:1978tb,Witten:1982fp}:
in which $H_y$ is slowly varying with respect to $y$.
In such cases, the Dirac equation $\cD\psi=0$ has an approximate solution
of the form $\psi=e^{-\int^y dy\,\lambda}\psi_\lambda$, where
$\psi_\lambda$ is an eigenfunction of $H_y$ with eigenvalue $\lambda(y)$.
This solution is normalizable when $\lambda>0$ and $\lambda<0$
as $y\ra+\infty$ and $y\ra-\infty$, respectively. Similarly, a
normalizable approximate solution of $\cD^\dag\psi=0$ is given by
$\psi=e^{+\int^y dy\,\lambda}\psi_\lambda$ with $\lambda<0$ and $\lambda>0$
as $y\ra+\infty$ and $y\ra-\infty$, respectively. Therefore,
the index is given by the difference of the number of eigenvalues that
cross zero from below and above when $y$ is increased from $y_-$ to $y_+$.
}
\begin{eqnarray}
\Ind(\cD|_I)
&=&\half\sum_{i=1}^k
\left(\eta(H_{y_i+\epsilon})-\eta(H_{y_i-\epsilon})\right)
\nn\\
&=&\half\left(\eta(H_{y_+})-\eta(H_{y_-})\right)
-\half\sum_{i=0}^k
\left(\eta(H_{y_{i+1}-\epsilon})-\eta(H_{y_i+\epsilon})\right)
\nn\\
&=&\half\left(\eta(H_{y_+})-\eta(H_{y_-})\right)
-\half\sum_{i=0}^k
\int_{y_i}^{y_{i+1}}dy\,\del_y\eta(H_y)\ ,
\label{indDI}
\end{eqnarray}
where $y_0\equiv y_-$ and $y_{k+1}\equiv y_+$.
Using (\ref{key2}) and (\ref{Iint}), we obtain
the desired result (\ref{APS2}):
\begin{eqnarray}
\Ind(\cD|_I)
=\half\left(\eta(H_{y_+})-\eta(H_{y_-})\right)
+\lim_{\Lambda\ra\infty}\int_{y_-<y<y_+}[\ch(\cF)]_{2r}\ .
\end{eqnarray}

Here, the boundary conditions for the fermions are such that the wave
function does not blow up at $y\ra\pm\infty$. In these regions, the
Dirac equation $\cD\psi=0$ with (\ref{cDHy}) can be solved by
\begin{eqnarray}
 \psi=e^{-\lambda_\pm y}\psi_{\lambda_\pm}\ ,
\end{eqnarray}
where $\psi_{\lambda_\pm}$ is an eigenfunction of $H_{y_\pm}$ with
the eigenvalue $\lambda_\pm$. Therefore, the modes with $\lambda_+<0$
and $\lambda_->0$ are discarded, which gives the APS boundary conditions.

Note that the formula (\ref{APS2}) is valid only for the
finite interval $I=[y_-,y_+]$. When, one wish to apply it for the
cases with $y_-\ra -\infty$ and/or $y_+\ra+\infty$, one should be
careful about the order of the limit $y_\pm\ra\pm\infty$ and
$\Lambda\ra\infty$, because they do not commute when
the mass diverges at $y\ra\pm\infty$, as we have seen in many
examples in section \ref{app}.
Let us consider a system defined on $M\times \R$ with
mass diverging at $y\ra\pm\infty$.
Suppose $|y_\pm|$ are
large enough so that $H_y$ does not have
a zero eigenvalue for any $y$ satisfying $y<y_-$ or $y_+<y$.
Then, (\ref{indDI}) implies that the index $\Ind(\cD|_I)$
is the same as that for $I=\R$.
Therefore, in this case, comparing (\ref{index})
and (\ref{APS2}), we obtain
\begin{eqnarray}
\half
\eta(H_{y_+})
-\lim_{\Lambda\ra\infty}
\int_{y_+<y}[\ch(\cF)]_{2r}
=
\half\eta(H_{y_-})
+\lim_{\Lambda\ra\infty}
\int_{y<y_-}[\ch(\cF)]_{2r}
\ .
\end{eqnarray}
Since the field configuration of the left hand side and
the right hand side are independent, we find
\begin{eqnarray}
\eta(H_{y_+})&=&
2\lim_{\Lambda\ra\infty}
\int_{y_+<y}[\ch(\cF)]_{2r}+\eta_0\ ,
\\
\eta(H_{y_-})&=&
-2\lim_{\Lambda\ra\infty}
\int_{y<y_-}[\ch(\cF)]_{2r}+\eta_0
\ ,
\end{eqnarray}
with a field-independent constant $\eta_0$.
Using these relations, we obtain
\begin{eqnarray}
\Ind(\cD|_{[y_-,+\infty]})
&=&\half\left(\eta_0-\eta(H_{y_-})\right)
+\lim_{\Lambda\ra\infty}\int_{y_-<y}[\ch(\cF)]_{2r}\ ,
\label{indplus}
\\
\Ind(\cD|_{[-\infty,y_+]})
&=&\half\left(\eta(H_{y_+})-\eta_0\right)
+\lim_{\Lambda\ra\infty}\int_{y<y_+}[\ch(\cF)]_{2r}\ .
\label{indminus}
\end{eqnarray}
These formulas are formally the same as (\ref{APS2})
with $[y_-,y_+]$ replaced with $[y_-,+\infty]$ or $[-\infty,y_+]$,
and $\eta(H_{\pm\infty})$ replaced with $\eta_0$.
Note that the second term in the right hand side of (\ref{indplus})
and (\ref{indminus}) is the generalized (gauge invariant) CS-form
given in (\ref{omegachF}) integrated over $M$.

For example, let us consider the case with compact $M$.
As a simple field configuration, we choose $A_-=A_+=0$
and $m=uy1_N$ with a real non-zero constant $u$.
In this case, we have
\begin{eqnarray}
 H_y=\mat{-i\gamma^\mu\del_\mu,uy,uy,i\gamma^\mu\del_\mu}\ ,
~~~H_y^2=\mat{-\del^2+(uy)^2,0,0,-\del^2+(uy)^2}\ ,
\end{eqnarray}
and $\eta(H_y)$ is trivially zero for any $y\ne 0$.
This implies $\eta_0=0$.

\section{Consistent vs. covariant anomalies}
\label{app:concov}

For the massless cases, it is well-known that the consistent and
covariant anomalies are related by the Bardeen-Zumino
counterterm.\cite{Bardeen:1984pm}
In this appendix, we review the relation between consistent and
covariant anomalies, and sketch the derivation of the Bardeen-Zumino
counterterms for the cases with spacetime dependent mass
in the covariant anomaly for completeness. Our strategy is to find
a counterterm to be added to the covariant anomaly so that
it satisfies the Wess-Zumino consistency condition. Note, however,
that this approach is not powerful enough to fix the mass dependence
of the anomaly $(D+2)$-form for the consistent anomaly.
We also point out that anomalous violation
of current conservation laws can be written in terms of
supermatrix-valued currents.

\subsection{Wess-Zumino consistency condition}
\label{WZcon}

Let us first introduce the notations for the consistent and covariant
anomalies as
\begin{eqnarray}
G(v)&\equiv& \delta_v\Gamma[A,m]\ ,
\label{GdeltaGam}
\\
\Gcov(v)
&\equiv& \int_M I_D^{1\,\textrm{cov}}(v,A,\wt m)\ ,
\end{eqnarray}
respectively, where $\Gamma[A,m]$ is the effective action
defined in (\ref{pf}),
$M$ is the $D$-dimensional spacetime
and $I_D^{1\,\textrm{cov}} $ is given in (\ref{covsuper})
and (\ref{covsuperD}).
By definition, the consistent anomaly $G(v)$ satisfies
the Wess-Zumino consistency condition \cite{Wess:1971yu}
\begin{eqnarray}
 \delta_{v_1}G(v_2)-\delta_{v_2}G(v_1)= G([v_1,v_2])\ .
\label{WZ}
\end{eqnarray}
On the other hand, it is easy to check from the explicit expression
that the covariant anomaly satisfies
\begin{eqnarray}
\delta_{v_1}\Gcov(v_2)=\Gcov([v_1,v_2])\ ,
\end{eqnarray}
which implies
\begin{eqnarray}
 \delta_{v_1}\Gcov(v_2)-\delta_{v_2}\Gcov(v_1)= 2\Gcov([v_1,v_2])\ ,
\end{eqnarray}
and hence the Wess-Zumino consistency condition is not satisfied.

The claim is that $G(v)$ and $\Gcov(v)$ are related (up to surface
terms and the gauge variation of local counterterms) by
\begin{eqnarray}
G(v)=\Gcov(v)+\alpha(v)
\label{concov}
\end{eqnarray}
with
\begin{eqnarray}
\alpha(v)\equiv
\left(\frac{i}{2\pi}\right)^{D/2}
\int_M\int_0^1 dt\,t\,\left[\Strsym\left(\sD v\,
e^{td\cA+t^2\cA^2}\cA\right)\right]_D\ ,
\label{alpha}
\end{eqnarray}
where $\cA$ is the superconnection (\ref{cA}) or (\ref{cA2})
for even or odd dimensions, respectively,
with $T=\wt m=m/\Lambda$ and
\begin{eqnarray}
\sD v\equiv dv+[\cA,v]=\delta_v\cA\ .
\label{sDv}
\end{eqnarray}
Here, $\Strsym$ denotes the symmetrized supertrace,
in which $\sD v$, $td\cA+t^2\cA^2$ and $\cA$ are symmetrized
(taking into account the sign flip when the odd elements
(such as $\sD v$ and $\cA$) are exchanged) before taking the supertrace.

Let us show that the right hand side of (\ref{concov})
satisfies the Wess-Zumino consistency condition
(\ref{WZ}).
For this purpose, it is convenient to rewrite $\alpha(v)$ as
\begin{eqnarray}
 \alpha(v)=
-\left(\frac{i}{2\pi}\right)^{D/2}
\int_{M\times I}\left[\Str\left(\delta_v\wt\cA\,
e^{\wt\cF}\right)\right]_{D+1}\ ,
\end{eqnarray}
where $I\equiv [0,1]\ni t$ and
\begin{eqnarray}
\wt\cA\equiv t\cA\ ,~~~
\wt\cF\equiv \wt d\wt\cA+\wt\cA^2
=td\cA+t^2\cA^2+dt\cA
\ ,~~~
\wt d\equiv d+dt\frac{\del}{\del t}\ .
\end{eqnarray}
We also define covariant derivatives $\sD$ and $\wt\sD$ as
\begin{eqnarray}
\sD\eta\equiv d\eta+\cA\,\eta-(-1)^{|\eta|}\eta\,\cA\ ,~~~
\wt\sD\wt\eta\equiv\wt d\wt\eta+\wt\cA\,\wt\eta-(-1)^{|\wt\eta|}
\wt\eta\,\wt\cA\ ,
\end{eqnarray}
where $\eta$ and $\wt\eta$ are supermatrix-valued fields
in $M$ and $M\times I$, respectively, and $|\eta|$ and
$|\wt\eta|$ denote their fermion numbers (mod 2).\footnote{
Recall that the differential form $dx^\mu$
and $\sigma^\pm$ are treated as fermions. See section \ref{review}.}

Using the relations
\begin{eqnarray}
&&\delta_{v_1} \delta_{v_2}\cA-\delta_{v_1} \delta_{v_2}\cA
=\delta_{[v_1,v_2]}\cA\ ,\\
&&\delta_v\wt\cF
=\wt d\delta_v\wt\cA+\wt\cA\delta_v\wt\cA+\delta_v\wt\cA\wt\cA
=\wt\sD\delta_v\wt\cA\ ,
\end{eqnarray}
and the Bianchi identity
\begin{eqnarray}
\wt\sD\wt\cF=\wt d\wt\cF+\wt\cA\wt\cF-\wt\cF\wt\cA=0\ ,
\end{eqnarray}
One can show
\begin{eqnarray}
\delta_{v_1}\alpha(v_2)-\delta_{v_2}\alpha(v_1)-\alpha([v_1,v_2])
&=&
-\left(\frac{i}{2\pi}\right)^{D/2}\int_{M\times I}
\Strsym\left(\wt\sD\left(
\delta_{v_1}\wt\cA\,\delta_{v_2}\wt\cA\,e^{\wt\cF}\right)
\right)\nn\\
&=&
-\left(\frac{i}{2\pi}\right)^{D/2}\int_{M\times I}
\wt d\,\Strsym\left(\delta_{v_1}\wt\cA\,\delta_{v_2}\wt\cA\,e^{\wt\cF}
\right)\ .
\nn\\
\label{mxI}
\end{eqnarray}
Using Stokes' theorem and
dropping the surface terms on the boundary of $M$,\footnote{
We only keep the parts that contribute to the
anomaly $(D+2)$-form for the consistent anomaly.}
 the right hand side of (\ref{mxI}) is evaluated as
\begin{eqnarray}
\int_{M\times I}
\wt d\,\Strsym\left(\delta_{v_1}\wt\cA\,\delta_{v_2}\wt\cA\,e^{\wt\cF}
\right)
&=&
\int_{M}\Strsym\left(\delta_{v_1}\cA\,\delta_{v_2}\cA\,e^{\cF}
\right)\nn\\
&=&\int_{M}\Strsym\left(\sD v_1\sD v_2\,e^{\cF}
\right)\nn\\
&=&\int_{M}\left( d\, \Strsym\left(v_1\sD v_2\,e^{\cF}\right)
-\Strsym\left(v_1\sD^2v_2\,e^{\cF}\right)\right)\nn\\
&=&\int_{M}
\Strsym\left(v_1[v_2,\cF]\,e^{\cF}\right)\nn\\
&=&\int_{M}
\Str\left([v_1,v_2]\,e^{\cF}\right)\ ,
\end{eqnarray}
where we have used
\begin{eqnarray}
D\cF=d\cF+\cA\cF-\cF\cA=0\ ,~~~
\sD^2v=d\sD v+\cA\sD v+\sD v\cA=
[\cF,v]\ .
\end{eqnarray}
Therefore, we get
\begin{eqnarray}
 \delta_{v_1}\alpha(v_2)-\delta_{v_2}\alpha(v_1)-\alpha([v_1,v_2])
=-\Gcov([v_1,v_2])\ ,
\end{eqnarray}
which implies that the right hand side of (\ref{concov})
satisfies the Wess-Zumino consistency condition (\ref{WZ}).

In section \ref{ID+2}, we used the fact that there is no difference
between the consistent and covariant anomalies for the $U(1)_V$
transformation when the background $U(1)_V$ gauge field $V$ is turned off.
This fact can be easily seen from the expression of $\alpha(v)$
in (\ref{alpha}). When $v$ is proportional to the unit matrix
and the $U(1)_V$ gauge field $V$ is set to zero, $\alpha(v)$ in (\ref{alpha})
can be written as
\begin{eqnarray}
\alpha(v)=\int_M\delta_vV \beta(\cA_0)
=\int_M\delta_v(V \beta(\cA_0))\ ,
\label{alphabeta}
\end{eqnarray}
where $\cA_0\equiv \cA|_{V=0}$ and
\begin{eqnarray}
\beta(\cA_0)\equiv\left(\frac{i}{2\pi}\right)^{D/2}
\int_0^1 dt\,t\,\left[\Strsym\left(
e^{td\cA_0+t^2\cA_0^2}\cA_0\right)\right]_{D-1}\ .
\end{eqnarray}
Therefore, this part can be canceled by
the gauge variation of a local counterterm.

\subsection{Currents and the Bardeen-Zumino counterterm}

The gauge variation of the effective action $\Gamma[A,m]$
can be written as
\begin{eqnarray}
\delta_v \Gamma[A,m]
=\int d^D x\,\left((\sD_\mu v)^a J^\mu_a
+(\sD v)^\alpha J_\alpha\right)\ ,
\label{deltaGamma}
\end{eqnarray}
where
\begin{eqnarray}
J^\mu_a(x)\equiv\frac{\delta\Gamma[A,m]}{\delta A_\mu^a(x)}\ ,~~~
J_\alpha(x)\equiv\frac{\delta\Gamma[A,m]}{\delta \wt m^\alpha(x)}\ .
\label{JJ}
\end{eqnarray}
Here, $A_\mu^a$ and $\wt m^\alpha=m^\alpha/\Lambda$ are the components
of the gauge field and the mass rescaled by a constant $\Lambda$,
and  $(\sD_\mu v)^a=(\delta_v A_\mu)^a$ and
$(\sD v)^\alpha=(\delta_v\wt m)^\alpha$ are their infinitesimal
gauge variations. (See (\ref{sDv}).)
$J^\mu_a$ and $J_\alpha$ in (\ref{JJ}) are the vacuum expectation values
of the currents $\delta S/\delta A_a^\mu$
and the fermion bilinear operators $\delta S/\delta \wt m^\alpha$,
respectively. Note that $\Lambda$ here is just an arbitrary parameter.
In fact, (\ref{deltaGamma}) does not depend on $\Lambda$.

$J^\mu_a$ and $J_\alpha$ can be considered as components of a
supermatrix-valued current analogous to the superconnection (\ref{cA}).
To see this explicitly, we choose a basis of the supermatrices
$\{T_a,T_\alpha\}$ such that the superconnection can be written as
$\cA=A^a_\mu dx^\mu T_a+\wt m^\alpha T_\alpha$
and introduce a dual basis $\{T^a,T^\alpha\}$ satisfying
\begin{eqnarray}
\Str(T_aT^b)=\delta_a^b\ ,~~~
\Str(T_\alpha T^\beta)=\delta_\alpha^\beta\ ,~~~
\Str(T_aT^\beta)=0\ ,~~~
\Str(T_\alpha T^b)=0\ .
\end{eqnarray}
A supermatrix-valued current is defined as
\begin{eqnarray}
\sJ(x)\equiv *J^{(1)}_a(x)\, T^a+*J^{(0)}_\alpha(x)\, T^\alpha\ ,
\label{sJ}
\end{eqnarray}
where $*$ is the Hodge star operator:
\begin{eqnarray}
*J^{(1)}_a(x)&\equiv&
\frac{1}{(D-1)!} \epsilon_{\mu_1\cdots\mu_D}J_a^{\mu_1}(x)\,
dx^{\mu_2}\cdots dx^{\mu_D}\ ,
\\
*J^{(0)}_\alpha(x)&\equiv&
J_\alpha(x)\, dx^1\cdots dx^D\ .
\end{eqnarray}
Using this, (\ref{deltaGamma}) can be written as
\begin{eqnarray}
 \delta_v\Gamma[A,m]=\int\Str(\sD v\,\sJ)\ ,
\end{eqnarray}
and the anomaly equation, obtained as the functional derivative of
(\ref{GdeltaGam}) with respect to $v(x)$, becomes
\begin{eqnarray}
*(\sD\sJ)_a=-\frac{\delta G(v)}{\delta v^a}\ ,
\label{DJ}
\end{eqnarray}
which shows that the consistent anomaly $G(v)$ represents the anomalous
violation of the current conservation law.
For example, for the axial $U(1)$ symmetry
(with $v_+=-v_-=-i\alpha 1_N$) in 4-dimensions,
the left hand side of (\ref{DJ}) becomes
\begin{eqnarray}
*(\sD\sJ)_{U(1)_A}
= \del_\mu\VEV{\ol\psi\gamma^\mu\gamma^5\psi}
+2im\VEV{\ol\psi \gamma^5\psi}\ ,
\end{eqnarray}
and, together with the right hand side obtained from
(\ref{delGam2})\footnote{In this local expression without integration
over spacetime, the $\wt m$ dependence in (\ref{delGam2}) drops out
in the $\Lambda\ra\infty$ limit with fixed $m$.},
(\ref{DJ}) reduces to the well-known formula for the axial
$U(1)$ anomaly.

{}From the expression (\ref{alpha}), we find that
$\alpha(v)$ can be written in the form
\begin{eqnarray}
\alpha(v)=\int_M d^Dx
\left((\sD_\mu v)^a P^\mu_a+(\sD v)^\alpha P_\alpha\right)
=\int\Str(\sD v\,\sP)
\end{eqnarray}
where $P_a^\mu$ and $P_\alpha$ are local functions of the gauge field
and the mass, and $\sP\equiv *P_aT^a+*P_\alpha T^\alpha$.
Then, the relation (\ref{concov}) implies that the covariant anomaly
is understood as the anomalous violation of conservation laws
\begin{eqnarray}
*(\sD\sJ^{\textrm{cov}})_a=-\frac{\delta \Gcov(v)}{\delta v^a}\ ,
\end{eqnarray}
for the covariant currents defined by
\begin{eqnarray}
J^{\textrm{cov}\,\mu}_a(x)\equiv J^\mu_a(x)-P^\mu_a(x)\ ,~~
J^{\textrm{cov}}_\alpha(x)\equiv J_\alpha(x)-P_\alpha(x)\ ,~~
\sJ^{\textrm{cov}}(x)\equiv\sJ(x) -\sP(x)\ .
\nn\\
\label{JJ2}
\end{eqnarray}
These $P^\mu_a$, $P_\alpha$ and $\sP$ are the Bardeen-Zumino
counterterms generalized to include the space-time dependent mass.

\end{document}